\documentclass[english, 12pt]{article}
\usepackage[english]{babel,varioref}
\usepackage[ansinew]{inputenc}
\usepackage[fleqn,reqno]{amsmath}
\usepackage{paralist}
\usepackage{natbib}
\usepackage{arydshln}
\usepackage{setspace}
\usepackage{graphics}

\bibpunct[: ]{(}{)}{,}{a}{}{,}

 \setdefaultenum{(i)}{a)}{(1)}{(A)}

\def\min{\mathop{\rm min}}
\def\max{\mathop{\rm max}}

\begin{document}

\title{Randomization does not help much, comparability does \linebreak
}

%(Submitted to Synthese)
%
%Über die Statistik und ihre Bedeutung für Philosophie und
%Wissenschaft

%\author{Uwe Saint-Mont }

\maketitle

%\vspace{10ex}
%\begin{quote} There is nothing in the world so difficult as that task
%of making up one’s mind. Who is there that has not longed that the power and
%privilege of selection among alternatives should be taken away from him in some
%important crisis of his life, and that his conduct should be arranged for him,
%either this way or that, by some divine power if it were possible — by some
%patriarchal power in the absence of divinity — or by chance even, if nothing
%better than chance could be found to do it? But no one dares to cast the die,
%and to go honestly by the hazard. There must be the actual necessity of obeying
%the die, before even the die can be of any use.

%\citep[Phineas Finn, chapter 60]{tr67}\footnote{See also \citet[431]{li82} and
%\citet[427]{ha88}}
%\end{quote}

\begin{doublespace}

{\bf Abstract}

Following Fisher, it is widely believed that randomization ``relieves the
experimenter from the anxiety of considering innumerable causes by which the
data may be disturbed.'' In particular, it is said to control for known and
unknown nuisance factors that may considerably challenge the validity of a
result. Looking for quantitative advice, we study a number of straightforward,
mathematically simple models. However, they all demonstrate that the optimism
with respect to randomization is wishful thinking rather than based on fact. In
small to medium-sized samples, random allocation of units to treatments
typically {\it yields} a considerable imbalance between the groups, i.e.,
confounding due to randomization is the rule rather than the exception.

In the second part of this contribution, we extend the reasoning to a number of traditional arguments for and against randomization. This discussion is rather non-technical, and at times even ``foundational'' (Frequentist vs. Bayesian). However, its result turns out to be quite similar. While randomization's contribution remains questionable, comparability contributes much to a compelling conclusion. Summing up, classical experimentation based on sound background theory and the systematic construction of exchangeable groups seems to
be advisable.

\vspace{15ex}

{\bf Key Words}. Randomization, Comparability, Confounding, Experimentation,
Design of Experiments, Statistical Experiments

\vspace{3ex} {\bf AMS-Classification:} 62K99

\tableofcontents

\newpage
\section{The Logic of the Experiment}\label{logic}

Randomization, the allocation of subjects to experimental conditions via a
random procedure, was introduced by R.A. \citet{fi35}. Arguably, it has since
become the most important statistical technique. In particular, statistical
experiments are defined by the use of randomization \citep{ro02, sh02}, and many
applied fields, such as {\it evidence based medicine}, draw a basic distinction
between randomized and non-randomized evidence \citep[e.g. the][]{ox09}.

In order to explain randomization's eminent role, one needs to refer to the
logic of the experiment, largely based on J. S. Mill's (1843: 225) {\it method
of difference}:
\begin{quote}
If an instance in which the phenomenon under investigation occurs, and an
instance in which it does not occur, have every circumstance in common save one,
that one occurring only in the former: the circumstance in which alone the two
instances differ, is the effect, or cause, or a necessary part of the cause, of
the phenomenon.
\end{quote}

Therefore, if one compares two groups of subjects (Treatment $T$ versus Control $C$, say)
and observes a substantial difference in the end (e.g. ${\bar X}_T > {\bar X}_C $),
that difference must be due to the experimental manipulation - IF the groups were
equivalent at the very beginning of the experiment. In other words, since the
difference between treatment and control (i.e. the experimental manipulation) is the
only perceivable reason that can explain the variation in the observations, it must
be the cause of the observed effect (the difference in the end). The situation is
quite altered, however, if the two groups already differed substantially at the
beginning. Then, there are two possible explanations of an effect:

\vspace{1ex}
\begin{center}
\begin{tabular}{|c|ccc|ccc|}
  \hline
  % after \\: \hline or \cline{col1-col2} \cline{col3-col4} ...
  Start of Experiment & $T$ & $=$ & $C$ & $T$ & $\neq$ & $C$ \\\hdashline
Intervention & Yes &  & No & Yes &  & No \\\hdashline
End of Experiment &&&&&& \\
(Observed Effect) & ${\bar X}_T$ & $>$ & ${\bar X}_C $ & ${\bar X}_T$ & $>$ & ${\bar
X}_C $
\\\hline
 Conclusion & \multicolumn{3}{|c|}{Intervention}   & Intervention & OR & Prior Difference  \\
 &  \multicolumn{3}{|c|}{caused the effect} &  \multicolumn{3}{|c|}{between the groups caused the effect} \\
  \hline
\end{tabular}
\end{center}

\vspace{1ex}

\section{Comparability}

Thus, for the logic of the experiment, it is of paramount importance to ensure
equivalence of the groups at the beginning of the experiment. The groups, or even
the individuals involved, must not be systematically different; one has to compare
like with like. Alas, in the social sciences exact equality of units, e.g. human
individuals, cannot be maintained. Therefore one must settle for {\it comparable}
subjects or groups $(T \approx C)$.

In practice, it is straightforward to define comparability with respect to the
features or properties of the experimental units involved. In a typical experimental
setup, statistical units (e.g. persons) are represented by their
corresponding vectors of attributes (properties, variables) such as gender, body
height, age, etc.:\vspace{1ex}

\begin{center}
\begin{tabular}{|c|c|c|c|c|c|c|c|}
  \hline
  % after \\: \hline or \cline{col1-col2} \cline{col3-col4} ...
  Unit No. & Gender & Height & Marital status & Age & Income & $\ldots$ & Attribute $j$ \\\hline
  1 & f (1) & 170cm & married (1) & 34 & \pounds 50000 & $\ldots$ & $a_{1j}$ \\
  2 & m (2) & 188cm & divorced (2) & 44 & \pounds 70000 & $\ldots$ & $a_{2j}$ \\
  3 & m (2) & 169cm & single (0) & 44 & \pounds 35000 & $\ldots$ & $a_{3j}$ \\
  \ldots & \ldots & \ldots & \ldots & \ldots&  \ldots &  \ldots &  \ldots\\
  $i$ & $a_{i1}$ & $a_{i2}$ & $a_{i3}$ & $a_{i4}$ & $a_{i5}$ & \ldots & $a_{ij}$ \\
  \hline
\end{tabular}
\end{center}

\vspace{1ex} If the units are almost equal in as many properties as possible, they
should be comparable, i.e., the remaining differences shouldn't alter the
experimental outcome substantially. However, since, in general, vectors have to be
compared, there is not a single measure of similarity. Rather, there are quite a lot
of measures available, depending on the kind of data at hand. An easily
accessible and rather comprehensive overview may be found in

reference.wolfram.com/mathematica/guide/DistanceAndSimilarityMeasures.html

As an example, suppose a unit $i$ is represented by a binary vector ${\bf
a}_i=(a_{i1},\ldots,a_{im})$. The Hamming distance $d(\cdot,\cdot)$ between two such
vectors is the number of positions at which the corresponding symbols are different.
In other words, it is the minimum number of substitutions required to change one
vector into the other. Let ${\bf a}_1=(0,0,1,0)$, ${\bf a}_2=(1,1,1,0)$, and ${\bf
a}_3=(1,1,1,1)$. Therefore $d({\bf a}_1,{\bf a}_2)=2$, $d({\bf a}_1,{\bf a}_3)=3$,
$d({\bf a}_2,{\bf a}_3)=1$, and $d({\bf a}_i,{\bf a}_i)=0$. Having thus calculated a
reasonable number for the ``closeness'' of two experimental units, one next has to
consider what level of deviance from perfect equality may be tolerable.

Due to the reasons outlined above, coping with similarities is a tricky business. Typically many
properties (covariates) are involved and conscious (subjective) judgement seems to
be inevitable. That might at least partially explain why matching on the (rather
objective) scalar valued propensity score, which is the probability of being
assigned to $T$ given a set of observed properties, has become popular recently
\citep{ru06}.

An even more serious question concerns the fact that relevant factors may not have
been recorded or might be totally unknown. In the worst case, similarity with
respect to some known factors has been checked, but an unnoticed nuisance variable
is responsible for the difference between the outcome in the two groups.

Moreover, comparability depends on the phenomenon studied. A clearly visible
difference, such as gender, is likely to be important with respect to life
expectancy, and can influence some physiological and psychological variables such as
height or social behaviour, but it is independent of skin color or blood type. In
other words, experimental units do not need to be twins in any respect; it suffices
that they be similar with respect to the outcome variable under study.

Given a unique sample it is easy to think about a {\it reference set} of other
samples that are alike in all relevant respects to the one observed. However, even
Fisher failed to give these words a precise formal meaning \citep[see][]{jo88}.
Nowadays, epidemiologists use the term `unconfounded' in order to communicate the
same idea: ``$[\ldots]$ the effect of treatment is unconfounded if the treated and
untreated groups resemble each other in all relevant features'' \citep[196]{pe09}.
Pearl shows that this problem is tightly linked to Simpson's famous paradox
\citep{si51} and devotes a major part of his book to the development of a formal
calculus that is able to cope with it. Another major advance was the idea of {\it
exchangeability}, a concept proposed by \citet{fi74}:
\begin{quote}
$[\ldots]$ instead of judging whether two groups are similar, the investigator
is instructed to imagine a hypothetical {\it exchange} of the two groups (the
treated group becomes untreated, and vice versa) and then judge whether the
observed data under the swap would be distinguishable from the actual data.
\citep[196]{pe09}
\end{quote}
\citet{ba93} gives some history on this idea and suggests the term
`permutability' instead, ``which conveys the idea of replacing one thing by
another similar thing.''

\section{Experimental Techniques to Achieve Comparability}\label{compare}

There are a number of strategies to achieve comparability. Starting with the
experimental units, it is straightforward to match similar individuals, i.e., to
construct pairs of individuals that are alike in many (most) respects.
\citet[583]{bo53} says:
\begin{quote} You may match them individual for individual in respect to what
seems to be their most important determinable and presumably relevant
characteristics $[\ldots]$ You can match litter-mates in body-weight if your
subjects are animals, and you can advertise for twins when your subjects are
human.\end{quote}

An example could be

\begin{center}
\begin{tabular}{|c|c|c|c|c|c|c|} \hline
 & Gender & Height & Martial status & Age & Income & Academic
  \\\hline
  $T1$ & 1  & 170 & 1  & 34 & 50000 & 1 (yes) \\
  $C1$ & 1  & 165 & 1  & 30 & 55000 & 1 (yes) \\\hdashline
   $T2$ & 2 & 193 & 2 & 46 & 72000 & 0 (no) \\
  $C2$ & 2  & 188 & 2  & 44 & 70000 & 1 (yes) \\\hdashline
  \ldots & \ldots & \ldots & \ldots & \ldots & \ldots & \ldots\\
  \hline
\end{tabular}
\end{center}

\vspace{1ex} Looking at the group level ($T$ and $C$), another obvious strategy is
to balance all relevant variables when assigning units to groups. Many approaches of
this kind are discussed in \citet[140]{se00}, minimization being the most prominent
among them \citep{ta74, tr98}, as the latter authors explain:
\begin{quote}
In our study of aspirin versus placebo $[\ldots]$ we chose age, sex, operating
surgeon, number of coronary arteries affected, and left ventricular function. But in
trials in other diseases those chosen might be tumour type, disease stage, joint
mobility, pain score, or social class.

At the point when it is decided that a patient is definitely to enter a trial, these
factors are listed. The treatment allocation is then made, not purely by chance, but
by determining in which group inclusion of the patient would minimise any
differences in these factors. Thus, if group A has a higher average age and a
disproportionate number of smokers, other things being equal, the next elderly
smoker is likely to be allocated to group B. The allocation may rely on minimisation
alone, or still involve chance but “with the dice loaded” in favour of the
allocation which minimises the differences.
\end{quote}

%However:
%\begin{quote} There are many uncontrollable factors that enter into the getting
%of human stuff; human beings are usually resistent to an indiscriminate
%mixing-up and to that arbitrary selection combined with complete ignorance of
%the nature of the individuals involved which constitutes `chance selection'
%$[\ldots]$ anyone who has attempted to obtain `unselected' samples with human
%material knows what very careful selection is required to achieve this
%`unselected' state. \citep[337]{bo19}\end{quote}

%, und man sollte vielleicht Fishers
%pointierte Worte (siehe \citet{fi35}, p. 19) bei der Einführung der
%Randomisierung heutzutage nicht ganz so streng sehen. War es damals nötig, das
%innovative Verfahren der Randomisierung gegen die systematischen Verfahren zu
%etablieren,\footnote{siehe z. B. \citet{ke55}} so scheint es heute eher nötig zu
%sein, letztere gegen ein Übergewicht der ersteren in Schutz zu nehmen.

However, apart from being cumbersome and relying on the experimenter's expertise (in
particular in choosing and weighing the factors), these strategies are always open
to the criticism that unknown nuisance variables may have had a substantial impact
on the result. Therefore \citet[18f]{fi35} advised strongly against treating every
conceivable factor explicitly:
\begin{quote} $[\ldots]$ it would be impossible to present an exhaustive
list of such possible differences appropriate to any kind of experiment, because
the uncontrolled causes which may influence the result are always strictly
innumerable $[\ldots]$ whatever degree of care and experimental skill is
expended in equalising the conditions, other than the one under test, which are
liable to affect the result, this equalisation must always be to a greater or
less extent incomplete, and in many important practical cases will certainly be
grossly defective. \end{quote}

Instead, he made the best of it and put forward his arguably most famous
contribution: \begin{quote} The full procedure of randomization [is the method]
by which the validity of the test of significance may be guaranteed against
corruption by the causes of disturbance which have not been eliminated
$[\ldots]$ the random choice of the objects to be treated in different ways [is]
a complete guarantee of the validity of the test of significance
[$\ldots$Randomization] relieves the experimenter from the anxiety of
considering and estimating the magnitude of the innumerable causes by which the
data may be disturbed. \citep[19, 20, 44]{fi35}
\end{quote}
Consequently, ``randomization controls for all possible confounders, known and
unknown'' became a common slogan. \citet{tr98} explain: ``The primary objective of
randomisation is to ensure that all other factors that might influence the outcome
will be equally represented in the two groups, leaving the treatment under test as
the only dissimilarity.''
%However, it should be noted that randomization is just a prominent tool, not a the
%final goal in and of itself:
%\begin{quote} In statistics, the purpose of randomization is to achieve homogeneity in the
%sample units $[\ldots]$ it should be spelled out that stability and homogeneity are
%the foundation of the statistical solution, {\it not} the other way around. For
%instance, in a clinical trial, applications of a randomized study to new patients
%rely on both the stability and homogeneity assumptions of our biological systems.
%\citep[52, 70, emphasis in the original]{wa93}
%\end{quote}
In the same vein \citet[9f]{be05} says:
\begin{quote}
    The idea of randomization is to overlay a sequence of units (subjects, or
    patients) onto a sequence of treatment conditions. If neither sequence can
    influence the other, then there should be no bias in the assignment of the
    treatments, and the comparison groups should be comparable.
\end{quote}

In other words, the weight of evidence of a statistical experiment crucially
depends on how well the tool of randomization achieves its ultimate goal:
comparability. Non-comparable groups offer a straightforward alternative
explanation, undermining the logic of the experiment. Thus the common term {\it
randomized evidence} is a misnomer, if randomization fails to reliably yield
comparable groups.

\section{Randomization and Comparability}\label{randcomp}

Historically, Fisher's idea proved to be a great success. \citet[427]{ha88} states:
``Indeed randomization is so commonplace that anyone untroubled by the fundamental
principles of statistics must suppose that the practice is quite uncontroversial.''
Nevertheless, quite a few scientists have argued that randomization does not work as
advertised. They concluded:
\begin{enumerate}

\item \citet[125]{al85}: ``Randomised
allocation in a clinical trial does not guarantee that the treatment groups are
comparable with respect to baseline characteristics.''

\item \citet[266]{ur85}: ``It is a chilling thought that medical treatments that
are worthless may have been endorsed, and valuable ones discarded, after
randomized trials in which the treatment groups differed in ways that were known
to be relevant to the disease under study, but where the strict rules of
randomization were applied and no adjustment made.''

\item \citet{tr98}: ``Indeed, if there are many possible prognostic factors there
will almost certainly be differences between the groups despite the use of random
allocation. In a small clinical trial a large treatment effect is being sought, but
a large difference in one or more of the prognostic factors can occur purely by
chance. In a large clinical trial a small treatment effect is being sought, but
small but important differences between the groups in one or more of the prognostic
factors can occur by chance. $[\ldots]$ At this point the primary objective of
randomisation — exclusion of confounding factors — has failed.''

\item \citet[35]{gr99}: $[\ldots]$ random imbalances may be severe, especially if
the study size is small $[\ldots]$

\item \citet[21]{ro02}: ``The
statement that randomization tends to balance covariates is at best imprecise;
taken too literally, it is misleading $[\ldots]$ What is precisely true is that
random assignment of treatments can produce some imbalances by chance, but
common statistical methods, properly used, suffice to address the uncertainty
introduced by these chance imbalances.''

%\item \citet[516]{be04}: ``Yet some of the benefits ascribed to
%randomization, for example that it eliminates all selection bias $[\ldots]$ can
%better be described as fantasy than reality.''

\item \citet[9]{be05}: ``While it is certainly true that randomization is used for the
purpose of ensuring comparability between or among comparison groups, we will see
$[\ldots]$ that it is categorically not true that this goal is achieved.'' [See also
his figure (p. 32) entitled: ``Random Imbalance, No Selection Bias.'']

 \item \citet[94]{bo05}: ``Even with randomization the assumption of
exchangeability can be violated.''

%\item \citet[21]{he05}: %``$[\ldots]$ many statisticians $[\ldots]$ resolutely
%defend randomization. [However,] even under ideal conditions, unaided randomization
%cannot answer some very basic questions
%such as what fraction of a
%population benefits from a program. And in practice, contamination and cross
%over effects make randomization a far from sure-fire solution $[\ldots]$''
%``$[\ldots]$ the statistical literature converts a metaphor for outcome selection -
%randomization - into an ideal.''

 \item \citet[259]{ho06}: ``$[\ldots]$ {\it the chief concern when designing a
 clinical trial should be to make it unlikely that the experimental groups differ
 on factors that are likely to affect the outcome} $[\ldots]$ With this rule in mind,
 it is evident that a randomized allocation of subjects to treatments might sometimes
 be useful in clinical trials as a way of better balancing the experimental groups
 $[\ldots]$ {\it But randomized allocation is not absolutely necessary; it is no}
 sine qua non; {\it it is not the only or even always the best way of constructing the treatment groups
 in a clinical trial}'' (emphasis in the original).

\item \citet[465]{wo07}: ``It is entirely possible that any particular randomization
may have produced a division into experimental and control groups that is
unbalanced with respect to `unknown' factor $X [\ldots]$''

\item \citet[2039]{au08}: ``While randomization will, on average,
balance covariates between treated and untreated subjects, it need not do so in
any particular randomization.''

\item \citet{ch12}: ``Despite randomization,
imbalance in prognostic factors as a result of chance (chance imbalance) may
still arise, and with small to moderate sample sizes such imbalance
may be substantial.''
\end{enumerate}

Moreover, quite early, statisticians - in particular of the Bayesian persuasion - put
forward several rather diverse arguments against randomization \citep{sa62,
sa76, ru78, ba80, li82, ba88, ka90}. The latter authors discuss
the role of ``randomized designs as methodological insurance against a `biased' sample''. They conclude:
\begin{quote}
What is it concerning randomization that makes such a judgement (of irrelevance of
the allocation to the test outcomes) compelling for the reader? We can find none and
suspect that randomization has little to do with whatever grounds there are for the
belief that the allocation is irrelevant to the test results. \citep[335f]{ka90}
\end{quote}
Fortunately, it is not necessary at this point to delve into delicate philosophical matters
or the rather violent Bayesian-Frequentist debate (however, see Section \ref{defense}). Fairly elementary
probabilistic arguments suffice to demonstrate that the above criticism hits its
target: By its very nature a random mechanism provokes fluctuations in the
composition of $T$ and $C$, making these groups (rather often) non-comparable.
Therefore the subsequent argument has the advantage of being straightforward,
mathematical, and not primarily ``foundational''. Its flavour is Bayesian in the
sense that we are comparing the {\it actual groups} produced by randomization which
is the ``posterior view'' preferred by that school. At the same time its flavour is
Frequentist, since we are focusing on the properties of a certain random {\it
procedure} which is the ``design view'' preferred by this school.

There are not just two, but (at least) three, competing statistical philosophies: ``In
many ways the Bayesian and frequentist philosophies stand at opposite poles from
each other, with Fisher's ideas being somewhat of a compromise'' \citep[98]{ef98}.
Since randomization is a Fisherian proposal, a neutral quantitative analysis of his
approach seems to be appropriate, acceptable to all schools, and, in a sense, long
overdue. Down-to-earth physicist \citet[xxii]{ja03} explains:
\begin{quote}
One can argue with philosophy; it is not easy to argue with a computer printout,
which says to us: `Independently of all your philosophy, here are the facts of
actual performance.'
\end{quote}

To this end we resume our reasoning with a simple but striking example:

\subsection*{Greenland's example}

\citet[422]{gr90} came up with ``$[\ldots]$ the smallest possible controlled trial
$[\ldots]$ to illustrate one thing randomization does {\it not} do: It does {\it
not} prevent the epidemiologic bias known as confounding $[\ldots]$''(emphasis in
the original; for a related example involving an interaction see p.
\pageref{interaktion}). That is, he flips a coin once in order to assign two
patients to $T$ and $C$, respectively: If heads, the first patient is assigned to
$T$, and the second to $C$; if tails, the first patient is assigned to $C$, and the
second to $T$. Suppose ${\bar X}_T > {\bar X}_C $, what is the reason for the
observed effect? Due to the experimental design, there are two alternatives: either
the treatment condition differed from the control condition, or patient $P_1$ was
not comparable to patient $P_2$.

\vspace{1ex}
\begin{center}
\begin{tabular}{|ccc|c|ccc|}
  \hline
  % after \\: \hline or \cline{col1-col2} \cline{col3-col4} ...
   $P_1$ &  & $P_2$ & Start of Experiment & $P_2$ &  & $P_1$ \\\hdashline
 Yes &  & No & Intervention & Yes &  & No \\\hdashline  ${\bar X}_T$ & $>$
& ${\bar X}_C $ & End of Experiment & ${\bar X}_T$ & $>$ & ${\bar X}_C $
\\\hline
\end{tabular}
\end{center}

\vspace{1ex}

However, as each patient is only observed under either the treatment or control (the left hand
side or the right hand side of the above table), one cannot distinguish between the
patient's and the treatment's impact on the observed result. Therefore Greenland
concludes:
\begin{quote}
No matter what the outcome of randomization, the study will be completely
confounded.
\end{quote}
Suppose the patients are perfect twins with the exception of a single difference.
Then Greenland's example shows that randomization cannot even balance a {\it single}
nuisance factor. To remedy the defect, it is straightforward to increase $n$. But
how large an $n$ will assure comparability? Like many authors, Greenland states the
basic result in a {\it qualitative} manner: ``Using randomization, one can make the
probability of severe confounding as small as one likes by increasing the size of
the treatment cohorts'' \citep[423]{gr90}. However, no quantitative advice is given
here or elsewhere. Thus it
should be worthwhile studying a number of explicit and straightforward models, {\it
quantifying} the effects of randomization.

\section{Random Confounding}\label{models}

\subsection{Dichotomous factors}\label{dicho}

Suppose there is a nuisance factor $X$ taking the value 1 if present and 0 if
absent. One may think of $X$ as a genetic aberration, a medical condition, a psychological
disposition or a social habit. Assume that the factor occurs with probability
$p$ in a certain person (independent of anything else). Given this, $2n$ persons
are randomized into two groups of equal size by a chance mechanism independent
of $X$.

Let $S_1$ and $S_2$ count the number of persons with the trait in the first and
the second group respectively. $S_1$ and $S_2$ are independent random
variables, each having a binomial distribution with parameters $n$ and $p$. A
natural way to measure the extent of imbalance between the groups is
$D=S_1-S_2$. Obviously, $ED=0$ and $$\sigma^2(D)=\sigma^2(S_1)+
\sigma^2(-S_2)=2\sigma^2(S_1)=2np(1-p).$$

Iff $D =0$, the two groups are perfectly balanced with respect to factor $X$. In
the worst case $|D|=n$, that is, in one group all units possess the
characteristic, whereas it is completely absent in the other. For fixed $n$, let
the two groups be comparable if $|D| \le n/i$ with some $i \in \{1,\ldots, n
\}$. Iff $i=1$, the groups will always be considered comparable. However, the
larger $i$, the smaller the number of cases we classify as comparable. In
general, $n/i$ defines a proportion of the range of $|D|$ that seems to be
acceptable. Since $n/i$ is a positive number, and $S_1=S_2 \Leftrightarrow
|D|=0$, the set of comparable groups is never empty.

Given some constant $i(<n)$, the value $n/i$ grows at a linear rate in $n$, whereas
$\sigma(D)=\sqrt{2np(1-p)}$ grows much more slowly. Due to continuity, there is a
single point $n(i,k)$, where the line intersects with $k$ times the standard
deviation of $D$. Beyond this point, i.e. for all $n\ge n(i,k)$, at least as many
realizations of $|D|$ will be within the acceptable range $[0, n/i]$.
Straightforward algebra gives,
$$
 n_p(i,k) = 2p(1-p) i^2 k^2 .
$$

\vspace{2ex} {\bf Examples}

A typical choice could be $i=10$ and $k=3$, which specifies the requirement that
most samples be located within a rather tight acceptable range. In this case, one
has to consider the functions $n/10$ and $3\sqrt{2p(1-p)n}$. These functions of $n$
are shown in the following graph:

\begin{figure}[h]

\includegraphics{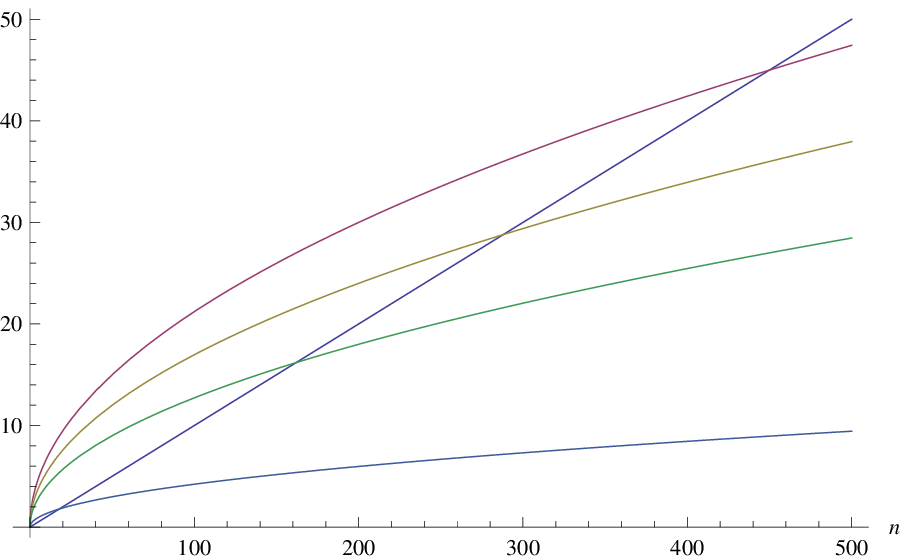}

Illustration: The linear function $n/10$, and $3\sqrt{2p(1-p)n}$ for $p=1/2, p=1/5, p=1/10,$ and $p=1/100$ (from above to below).
\end{figure}

Thus, depending on $p$, the following numbers of subjects are needed per group (and
twice this number altogether):
$$
\begin{array}{lll|l|l}
  p & i & k & n_p(i,k) & 2 \cdot n_p(i,k) \\\hline
  1/2 & 10 & 3 & 450 & 900 \\
  1/5 & 10 & 3 & 288 & 576 \\
  1/10 & 10 & 3 & 162 & 324 \\
  1/100 & 10 & 3 & 18 & 36 \
\end{array}
$$
Relaxing the criterion of comparability (i.e. a smaller value of $i$) decreases the
number of subjects necessary:
$$
\begin{array}{lll|l}
  p & i & k & n_p(i,k) \\\hline
  1/2 & 5 & 3 & 113 \\
  1/5 & 5 & 3 & 72 \\
  1/10 & 5 & 3 & 41 \\
  1/100 & 5 & 3 & 5 \
\end{array}
$$
The same happens if one decreases the number of standard
deviations $k$:%\hspace{10ex}
$$\begin{array}{lll|l}
  p & i & k & n_p(i,k) \\\hline
  1/2 & 10 & 2 & 200 \\
  1/5 & 10 & 2 & 128 \\
  1/10 & 10 & 2 & 72 \\
  1/100 & 10 & 2 & 8 \
\end{array}
$$

%For the sake of illustration the third line of the last tables (where $p=1/10$ and
%thus $2p(1-p)=9/50$) is given in the following graph, i.e. the functions $n/10, n/5,
%2\sqrt{9n/50}$, and $3\sqrt{9n/50}$ are displayed:

%\pagebreak

%\begin{figure}[h]

%\includegraphics{bild2}

%\end{figure}

This shows that randomization works, if the number of subjects ranges in the
hundreds or if the probability $p$ is rather low. (By symmetry, the same conclusion
holds if $p$ is close to one.) Otherwise there is hardly any guarantee that the two
groups will be comparable. Rather, they will differ considerably due to random
fluctuations.

The distribution of $D$ is well known \citep[e.g.][142f]{jo05}. For
$d=-n,\ldots,n,$
$$ P(D=d) = \sum_{\max(0,d) \le y \le \min(n,n+d)} \binom{n}{y} \binom{n}{y-d}
p^{2y-d} (1-p)^{2n-2y+d}
$$

Therefore, it is also possible to compute the probability $q=q(i, n, p)$ that
two groups, constructed by randomization, will be comparable.
%Obviously, this probability depends on $p$, $i$, and $n$.
If $i=5$, i.e., if one fifth of the range of $|D|$ is judged to be comparable, we
obtain:
$$
\begin{array}{ll|l}
  p &  n & q(i,n,p) \\\hline
  1/2  & 5 & 0.66 \\
  1/2  & 10 & 0.74 \\
  1/2  & 25 & 0.88 \\
  1/2  & 50 & 0.96 \
\end{array}
\hspace{5ex}
\begin{array}{ll|l}
  p & n & q(i,n,p) \\\hline
  1/10 & 5 & 0.898 \\
  1/10 & 10 & 0.94 \\
  1/10 & 25 & 0.98999 \\
  1/10 & 50 & 0.999 \
\end{array}
\hspace{5ex}
\begin{array}{ll|l}
  p & n & q(i,n,p) \\\hline
  1/100 & 5 & 0.998 \\
  1/100 & 10 & 0.9997 \\
  1/100 & 25 & 0.999999 \\
  1/100 & 50 & 1 \
\end{array}
$$

Thus, it is rather difficult to control a factor that has a probability of about
$1/2$ in the population. However, even if the probability of occurrence is only
about $1/10$, one needs more than 25 people per group to have reasonable
confidence that the factor has not produced a substantial imbalance.

\vspace{2ex} {\bf Several factors}

The situation becomes worse if one takes more than one nuisance factor into
account. Given $m$ independent binary factors, each of them occurring with
probability $p$, the probability that the groups will be balanced with respect
to all nuisance variables is $q^m$. Numerically, the above results yield:
$$
\begin{array}{lll|lll}
  p & n & q & q^2 & q^5 & q^{10} \\\hline
  1/2 & 5 & 0.66 & 0.43 & 0.12 & 0.015 \\
  1/2 & 10 & 0.74 & 0.54 & 0.217 & 0.047 \\
  1/2 & 25 & 0.88 & 0.78 & 0.53 & 0.28  \\
  1/2 & 50 & 0.96 & 0.93 & 0.84 & 0.699 \\
\end{array}
\hspace{5ex}
\begin{array}{lll|lll}
  p & n & q & q^2 & q^5 & q^{10} \\\hline
  1/10 & 5 & 0.898 & 0.807 & 0.58 & 0.34  \\
  1/10 & 10 & 0.94 & 0.88 & 0.74 & 0.54 \\
  1/10 & 25 & 0.98999 & 0.98 & 0.95 & 0.90 \\
  1/10 & 50 & 0.999 & 0.9989 & 0.997 & 0.995 \\
\end{array}
$$$$
\begin{array}{lll|lll}
  p & n & q & q^2 & q^5 & q^{10} \\\hline
  1/100 & 5 & 0.998 & 0.996 &  0.99 & 0.98 \\
  1/100 & 10 & 0.9998 & 0.9996 & 0.999 & 0.9979 \\
  1/100 & 25 & 0.9999998 & 0.9999995 & 0.999999 & 0.999998 \\
  1/100 & 50 & 1 & 1 & 1 & 1\\
\end{array}
$$

Accordingly, given $m$ independent binary factors, each occurring with probability
$p_j$ (and corresponding $q_j=q(i,n,p_j)$), the probabilities closest to $1/2$ will
dominate $1-q_1\cdots q_m$, which is the probability that the two groups are not
comparable due to an imbalance in at least one variable. In a typical study with
$2n=100$ persons, for example, it does not matter if there are one, two, five or
even ten factors, if each of them occurs with probability of $1/100$. However, if
some of the factors are rather common (e.g. $1/5 < p_j <4/5$), this changes
considerably. In a smaller study with fewer than $2n=50$ participants, a few such
factors suffice to increase the probability that the groups constructed by
randomization won't be comparable to 50\%. With only a few units per group, one can
be reasonably sure that some undetected, but rather common, nuisance factor(s) will
make the groups non-comparable.

The situation deteriorates considerably if there are interactions between the
variables that may yield convincing alternative explanations for an observed effect.
It is possible that all factors considered in isolation are reasonably balanced
(which is often checked in practice), but that a certain combination of them affects
the observed treatment effect. For the purpose of illustration suppose four persons
(being young or old, and male or female) are investigated:

%\vspace{1ex}
\begin{center}
\begin{tabular}{c|c}\label{interaktion}
  T & C \\\hline
  % after \\: \hline or \cline{col1-col2} \cline{col3-col4} ...
  Old Man & Old Woman  \\
 Young Woman & Young Man
\\\hline
%& \\
%${\bar X}_T$ & ${\bar X}_C$ \\
\end{tabular}
\end{center}

\vspace{1ex}

Although gender and (dichotomized) age are perfectly balanced between $T$ and $C$,
the young woman has been allocated to the first group. Therefore a property of young
women (e.g. pregnancy) may serve as an explanation for an observed effect, e.g.
${\bar X}_T > {\bar X}_C$.

Given $m$ factors, there are $m (m-1) /2$ possible interactions between just two of
the factors, and $\binom{m}{\nu}$ possible interactions between $\nu$ of them. Thus,
there is a high probability that some considerable imbalance occurs in at least one
of these numerous interactions, in small groups in particular. For a striking early
numerical study see \citet{le80}.

Detected or undetected, such imbalances provide excellent alternative explanations
of an observed effect. Altogether our conclusion based on an explicit quantitative
analysis coincides with the qualitative argument given by \citet[91]{sa62}:
\begin{quote}
    Suppose we had, say, thirty fur-bearing animals of which some were junior and
    some senior, some black and some brown, some fat and some thin, some of one
    variety and some of another, some born wild and some in captivity, some sluggish
    and some energetic, and some long-haired and some short-haired. It might be hard
    to base a convincing assay of a pelt-conditioning vitamin on an experiment with
    these animals, for every subset of fifteen might well contain nearly all of the
    animals from one side or another of one of the important dichotomies $[\ldots]$

    Thus contrary to what I think I was taught, and certainly used to believe, it
    does not seem possible to base a meaningful experiment on a small heterogenous
    group.
\end{quote}

In the light of this, one can only hope for some `benign' dependence structure among
the factors, i.e., a reasonable balance in one factor improving the balance in (some
of) the others. Given such a tendency, a larger number of nuisance factors may be
controlled, since it suffices to focus on only a few. Independent variables possess
a `neutral' dependence structure in that the balance in one factor does not
influence the balance in others. Yet, there may be a `malign' dependence structure,
such that balancing one factor tends to actuate imbalances in others.
%Given this, it is almost impossible to achieve comparability via randomization
%or any other method.

We will make this argument more precise in Section \ref{intermediate}. However, a
concrete example will illustrate the idea: Given a benign dependence structure,
catching one cow (balancing one factor) makes it easier to catch others. Therefore it is
easy to lead a herd into an enclosure: Grab some of the animals by their horns
(balance some of the factors) and the others will follow. However, in the case of a
malign dependence structure the same procedure tends to stir up the animals, i.e.,
the more cows are caught (the more factors are being balanced), the less
controllable the remaining herd becomes.

\subsection{Ordered random variables}

In order to show that our conclusions do not depend on some specific model, let us
next consider ordered random variables. To begin with, look at four units with ranks
1 to 4. If they are split into two groups of equal size, such that the best (1) and
the worst (4) are in one group, and (2) and (3) are in the other, both groups have
the same rank sum and are thus comparable. However, if the best and the second best
constitute one group and the third and the fourth the other group, their rank sums (3
versus 7) differ by the maximum amount possible, and they do not seem to be
comparable. If the units with ranks 1 and 3 are in the first group and the units
with ranks 2 and 4 are in the second one, the difference in rank sums is $|6-4|=2$
and it seems to be a matter of personal judgement whether or not one thinks of them
as comparable.

Given two groups, each having $n$ members, the total sum of ranks is
$r=2n(2n+1)/2=n(2n+1)$. If, in total analogy to the last section, $S_1$ and
$S_2$ are the sum of the ranks in the first and the second group, respectively,
$S_2=r-S_1$. Therefore it suffices to consider $S_1$, which is the test statistic of
Wilcoxon's test. Again, a natural way to measure the extent of imbalance between
the groups is $D=S_1-S_2=2S_1-r$. Like before $ED=0$ and because $\sigma^2(S_1)=
n^2(2n+1)/12$ we have $\sigma^2(D)=4\sigma^2(S_1)=n^2 (2n+1)/3$.

Moreover, $n(n+1)/2 \le S_j \le n(3n+1)/2$ $(j=1,2)$ yields $-n^2 \le D \le
n^2$. Thus, in this case, $n^2/i$ $(i \in \{1,\ldots,n^2 \})$ determines a
proportion of the range of $|D|$ that may be used to define comparability. Given
a fixed $i (<n^2)$, the quantity $n^2/i$ is growing at a quadratic rate in $n$,
whereas $\sigma(D)=n\sqrt{(2n+1)/3}$ is growing at a slower pace. Like before,
there is a single point $n(i,k)$, where $n^2/i = k \sigma(D)$. Straightforward
algebra gives,
$$
 n(i,k) = ik(ik+\sqrt{(ik)^2 +3})/3.
$$
Again, we see that large numbers of observations are needed to ensure
comparability:
$$\begin{array}{ll|l}
  i & k & n(i,k) \\\hline
  10 & 3 & 501 \\
  5 & 3 & 156 \\
  10 & 2 & 267 \\
\end{array}
$$

%The last two lines of this table can be found in the next graph, where the functions
%$n^2/5, n^2/10, 3 n \sqrt{(2n+1)/3}$, and $2 n \sqrt{(2n+1)/3}$ are displayed:

%\begin{figure}[h]

%\includegraphics{bild3}

%\end{figure}

Again, it is possible to work with the distribution of $D$ explicitly. That is,
given $i$ and $n$, one may calculate the probability $q=q(i, n)$ that two
groups, constructed by randomization, are comparable. If $|D| \le n^2 /i$ is
considered comparable, it is possible to obtain, using the function pwilcox() in
R:
$$\begin{array}{l|lllll}
 & \multicolumn{5}{c}{n}\\
 i &  5 & 10 & 25 & 50 & 100 \\\hline
  3 & 0.58 & 0.78 & 0.96 & 0.996 & 1 \\
  5 & 0.45 & 0.56 & 0.78 & 0.92 & 0.99 \\
  10 & 0.16 & 0.32 & 0.45 & 0.61 & 0.78 \\
\end{array}
$$

These results for ordered random variables are perfectly in line with the
conclusions drawn from the binary model. Moreover, the same argument as before shows that the
situation becomes (considerably) worse if several factors may influence the final
result.

\subsection{A continuous model}

Finally, I consider a continuous model. Suppose there is just one factor $X \sim
N(\mu,\sigma)$. One may think of $X$ as a normally distributed personal ability,
person $i$ having individual ability $x_i$. As before, assume that $2n$ persons
are randomized into two groups of equal size by a chance mechanism independent
of the persons' abilities.

Suppose that also in this model $S_1$ and $S_2$ measure the total amount of ability
in the first and the second group respectively. Obviously, $S_1$ and $S_2$ are
independent random variables, each having a normal distribution
$N(n\mu,\sqrt{n} \sigma)$. A straightforward way to measure the {\it absolute}
extent of imbalance between the groups is

\begin{equation}\label{gleich}
D=S_1-S_2=\sum_{\iota=1}^n X_{1,\iota}-\sum_{\iota=1}^n X_{2,\iota}=\sum_{\iota=1}^n
(X_{1,\iota}-X_{2,\iota}).
\end{equation}

Due to independence, obviously $D \sim N(0,\sqrt{2n}\sigma)$.

Let the two groups be comparable if $|D| \le l \sigma$, i.e., if the difference
between the abilities assembled in the two groups does not differ by more than
$l$ standard deviations of the ability $X$ in a single unit. The larger $l$, the
more cases are classified as comparable. For every fixed $l$, $l \sigma$ is a
constant, whereas $\sigma(D)=\sqrt{2n} \sigma$ is growing slowly. Owing to
continuity, there is yet another single point $n(l)$, where $l \sigma =
\sigma(D)=\sqrt{2n} \sigma$. Straightforward algebra gives,
$$
 l \le \sqrt{2n} \Leftrightarrow 2 n \ge  l^2 .
$$
In particular, we have:
$$
\begin{array}{l|llllllllll}
  l & 1 & 2 & 3 & 5 & 10 \\\hline
  n & 1 & 2 & 5 & 13 & 50 \\
\end{array}
$$
In other words, the two groups become non-comparable very quickly. It is almost
impossible that two groups of 500 persons each, for example, could be close to one
another with respect to total (absolute) ability.

However, one may doubt if this measure of non-comparability really makes sense.
Given two teams with a hundred or more subjects, it does not seem to matter whether
the total ability in the first one is within a few standard deviations of the other.
Therefore it is reasonable to look at the {\it relative} advantage of group 1 with
respect to group 2, i.e. $ Q = D/n$. Why divide by $n$ and not by some other
function of $n$? First, due to equation (\ref{gleich}), exactly $n$ comparisons
$X_{1,\iota}-X_{2,\iota}$ have to be made. Second, since $$Q=\sum_{\iota =1}^n
X_{1,\iota}/n -\sum_{\iota =1}^n X_{2,\iota}/n = {\bar X}_T - {\bar X}_C,$$ $Q$ may
be interpreted in a natural way, i.e., being the difference between the typical
(mean) representative of group 1 (treatment) and the typical representative of group
2 (control). A straightforward calculation yields $Q \sim N(0,\sigma \sqrt{2/n})$.

Let the two groups be comparable if $|Q| \le l \sigma$. If one wants to be
reasonably sure (three standard deviations of $Q$) that comparability holds, we have
$ l\sigma \ge 3\sigma \sqrt{2/n} \Leftrightarrow n \ge 18/l^2 $. Thus, at least the
following numbers of subjects are required per group:
$$
\begin{array}{l|llllll}
  l & 5 & 2 & 1 & 1/2 & 1/4 & 1/8 \\\hline
  n & 1 & 5 & 18 & 72 & 288 & 1152 \\
\end{array}
$$
If one standard deviation is considered a large effect \citep{co88}, three dozen
subjects are needed to ensure that such an effect will not be produced by chance. To
avoid a small difference between the groups due to randomization (one quarter of a
standard deviation, say), the number of subjects needed goes into the hundreds.
%This
%can also be seen in the following graphic where the non-constant function is
%$3\sqrt{2/n}$:

%\begin{figure}[h]

%\includegraphics{bild4}

%\end{figure}

In general, if $k$ standard deviations of $Q$ are desired, we have,
$$ n
\ge  2 k^2 /l^2.
$$
Thus, for $k=1,2$ and $5$, the following numbers of subjects $n_k$ are required
in each group:
$$
\begin{array}{l|llllll}
  l & 5 & 2 & 1 & 1/2 & 1/4 & 1/8 \\\hline
  n_1 & 1 & 1 & 2 & 8 & 32 & 128 \\
  n_2 & 1 & 2 & 8 & 32 & 128 & 512 \\
  n_5 & 2 & 13 & 50 & 200 & 800 & 3200 \\
\end{array}
$$
These are just the results for one factor. As before, the situation
deteriorates considerably if one sets out to control several nuisance variables
by means of randomization.

\section{Intermediate Conclusions}\label{intermediate}

The above models have deliberately been kept as simple as possible. Their
results are straightforward and they agree: If $n$ is small, it is almost impossible
to control for a trait that occurs frequently at the individual level, or for a
larger number of confounders, via randomization. It is of paramount importance to
understand that random fluctuations lead to considerable differences between small
or medium-sized groups, making them very often non-comparable, thus undermining the
basic logic of experimentation. That is, `blind' randomization does not create
equivalent groups, but rather {\it provokes} imbalances and subsequent artifacts: Even in
larger samples one needs considerable luck to succeed in creating
equivalent groups; $p$ close to 0 or 1, a small number of nuisance factors $m$, or a
favourable dependence structure that balances all factors, including their relevant
interactions, if only some crucial factors are to be balanced by chance.

``Had the trial not used random assignment, had it instead assigned patients one at
a time to balance [some] covariates, then the balance might well have been better
[for those covariates], but there would be no basis for expecting other unmeasured
variables to be similarly balanced'' \citep[21]{ro02} seems to be the only argument
left in favour of randomization. Since randomization treats known and unknown factors
alike, it is quite an asset that one may thus infer from the observed to the
unobserved without further assumptions. However, this argument backfires immediately
since, for exactly the same reason, an imbalance in an observed variable cannot be
judged as harmless. Quite the contrary: An observed imbalance {\it hints at} further
undetectable imbalances in unobserved variables.

Moreover, treating known and unknown factors equivalently is cold comfort compared
to the considerable amount of imbalance evoked by randomization. Fisher's favourite
method always comes with the cost that it introduces additional variability, whereas
a systematic schema at least balances known factors. In subject areas haunted by
heterogeneity it seems intuitively right to deliberately work in favour of
comparability, and rather odd to introduce further variability.

\subsection*{Three types of dependence structures}

In order to sharpen these arguments, let us look at an observed factor $X$, an
unobserved factor $Y$, and their dependence structure in more detail. Without loss
of generality let all functions $d(\cdot)$ be positive in the following. Having
constructed two groups of equal size via randomization, suppose $d_R(X)={\bar X}_T
-{\bar X}_C >0$ is the observed difference between the groups with respect to
variable $X$. Using a systematic scheme instead, i.e., distributing the units among
$T$ and $C$ in a more balanced way, this may be reduced to $d_S(X)$. The crucial
question is how such a manipulation affects $d_R(Y)$, the balance between the groups
with respect to another variable.

A benign dependence structure may be characterized by $d_S(Y) < d_R(Y)$. In
other words, the effort of balancing $X$ pays off, since the increased
comparability in this variable carries over to $Y$. For example, given today's
occupational structures with women earning considerably less than men, balancing
for gender should also even out differences in income. If balancing in $X$ has no
effect on $Y$, $d_S(Y) \approx d_R(Y)$, no harm is done. For example, balancing
for gender should not affect the distribution of blood type in the observed
groups, since blood type is independent of gender. Only in the
pathological case when increasing the balance in $X$ has the opposite effect on $Y$, %(e.g. if $Y=1/X$)
one may face troubles. As an example, let there be four pairs $(x_1,y_1)=(1,4);
(x_2,y_2)=(2,2);(x_3,y_3)=(3,1)$; and $(x_4,y_4)=(4,3)$. Putting units 1 and 4
in one group, and units 2 and 3 in another, yields a perfect balance in the
first variable, but the worst imbalance possible in the second.

However, suppose $d(\cdot) < c$ where the constant (threshold) $c$ defines
comparability. Then, in the randomized case, the groups are comparable if both
$d_R(X)$ and $d_R(Y)$ are smaller than $c$. By construction, $d_S(X)\le d_R(X)<c$,
i.e., the systematically composed groups are also comparable with respect to $X$.
Given a malign dependence structure, $d_S(Y)$ increases. Yet $d_S(Y)<c$ may still
hold, since, in this case, the ``safety margin'' $c-d_R(Y)$ may prevent the
systematically constructed groups from becoming non-comparable with respect to
property $Y$. In large samples, $c-d_R(\cdot)$ is considerable for both variables.
Therefore, in most cases, consciously constructed samples will (still) be comparable.
Moreover, the whole argument easily extends to more than two factors.

%Notice, however, that in all these cases, minimization or a related
%non-randomized scheme have the same advantages. Given a benign dependence
%structure, consciously balancing known factors also tends to balance correlated,
%but unknown confounders. In particular, if $m$ is small or if the variables are
%strongly correlated, it pays off to isolate and explicitly control for a handful
%of obviously influential factors which is a crucial ingredient of classical
%experimentation.

In a nutshell, endeavouring to balance relevant variables pays off. A
conscious balancing schema equates known factors better than chance and may have
some positive effect on related, but unknown, variables. If the
balancing schema has no effect on an unknown factor, the latter is treated as if randomization
were interfering - i.e. in a completely nonsystematic, `neutral' way. Only if
there is a (very) malign dependence structure, when systematically balancing some variable
provokes (considerable) ``collateral damage'', might randomization be preferable.

This is where sample size comes in. In realistic situations with many unknown
nuisance factors, randomization only works if $n$ is (really) large. Yet if $n$ is
large, so are the ``safety margins'' in the variables, and even an unfortunate
dependence structure won't do any harm. If $n$ is smaller, the above models show that
systematic efforts, rather than randomization, may yield comparability. Given a
small number of units, both approaches only have a chance of succeeding if there are
hardly any unknown nuisance factors, or if there is a benign dependence structure,
i.e., if a balance in some variable (no matter how achieved) has a positive effect on
others. In particular, if the number of relevant nuisance factors and interactions
is small, it pays to isolate and control for a handful of obviously influential
variables, which is a crucial ingredient of experimentation in the classical natural
sciences. Our overall conclusion may thus be summarized in the following table:

\vspace{1ex}
\begin{center}
\begin{tabular}{c|c|c|l}
  Dependence structure & X (observed) & Y (unobserved) & Preferable procedure \\\hline
  % after \\: \hline or \cline{col1-col2} \cline{col3-col4} ...
Benign & $d_S(X) < d_R(X)$ & $d_S(Y) < d_R(Y)$ & Systematic allocation  \\
Neutral & $d_S(X) < d_R(X)$ & $d_S(Y) \approx d_R(Y)$ & Systematic allocation  \\
Malign & $d_S(X) < d_R(X)$ & $d_S(Y) >d_R(Y)$ & Rather systematic than
 \\
&  &  &  random allocation
\\\hline
\end{tabular}
\end{center}

\vspace{1ex}

%Although, alas, the latter involves less work, the above models show that it is
%futile to hope that chance will do a better job. (This could be a modern version
%of a German proverb that says ``luck is with the diligent.'')

%The latter can only be avoided if the number of subjects is increased
%drastically or if some systematic scheme like balancing (e.g. via a
%pre-specified minimization scheme) {\it restricts} chance's crucial role.

\section{The Frequentist Position}\label{defense}

There is yet another important, some would say outstanding, defense of randomization that we have omitted so
far.
According to this point of view the major ``$[\ldots]$ function of
randomization is to generate the sample space and hence provide the basis for
estimates of error and tests of significance $[\ldots]$'' \citep[419]{co76}.
In a statistical experiment one controls the random mechanism, thus the experimenter knows the
sample space and the distribution in question. This constructed and therefore ``valid'' framework keeps nuisance variables at bay and sound reasoning within the framework leads to correct results. Someone following this train of thought could therefore state - and several referees of this contribution indeed did so - that the above models underline the rather well-known fact that randomization can have difficulties in constructing similar groups (achieving exchangeablility/comparability, balancing covariates), in particular if $n$ is small. However, this goal is quite subordinate to the major goal of establishing a known distribution on which sound statistical conclusions can be based.

The latter view has been proposed and defended by Frequentist statisticians, in particular Fisher and Neyman. It once dominated the field of statistics and still has a stronghold in certain quarters, in particular medical statistics where {\it randomized} controlled trials have been the gold standard. In this section, we focus on the basic Frequentist viewpoint and some its major criticisms. Given this, several perspectives on randomization will be the thrust of the next section. Section \ref{cause} then provides the link to causation and Section \ref{history} gives a wider, historical perspective.

Traditionally, criticism of the Frequentist line of argument in general, and randomization in particular, has come from the Bayesian school of statistics.
While Frequentist statistics is much concerned with the way data is collected, focusing on the design of experiments, the corresponding sample space and sampling distribution, Bayesian statistics is rather concerned with the data actually obtained. Its focus is on learning from the(se) data - in particular with the help of Bayes' theorem - and the parameter space.

In a sense, both viewpoints are perfectly natural and not contradicting each other, so it may seem futile to decide which emphasis is more appropriate. However, the example of randomization shows that this cannot be the final word: For the pre-data view, randomization is essential, it constitutes the difference between a real statistical experiment and any kind of quasi-experiment. For the post-data view, however, randomization adds nothing to the information at hand, and is ancilliary or just a nuisance.

Consistently, \citet[238]{fr08a}\label{free33}, an outstanding Frequentist, says a bit more generally that there are ``{\it two styles of
inference}.
\begin{itemize}
  \item Randomization {\it provides} a {\it known} distribution for the assignment
  variables; statistical inferences are based on this distribution.
  \item Modeling {\it assumes} a distribution for the latent variables;\label{latent4} statistical
  inferences are based on that assumption.'' (Emphasis added)
\end{itemize}

      Yet \citet{ja03}, an outstanding Bayesian, devoting a section (16.4) of his book to pre- and post-data considerations, states (p. 500): ``As we have stressed repeatedly, virtually all real problems of scientific inference are concerned with post-data questions.''

The crucial and rather fundamental issue therefore becomes how far-reaching the conclusions of each of these styles of inference are. To pin down the differences, \citet[192]{ba99} gives a nice and important example:
\begin{quote}
On the one hand, the procedure of using the sample mean (or some other
measure) to estimate $\mu$ could be assessed in terms of how well we expect
it to behave; that is, in the light of different possible sets of data that
might be encountered. It will have some average characteristics that
express the precision we initially expect, i.e. before we take our data
$[\ldots]$

The alternative concept of final precision aims to express the
precision of an inference in the specific situation we are studying. Thus,
if we actually take our sample and find ${\bar x}=29.8,$ how are we to answer
the question `how close is $29.8$ to $\mu$'? This is a most pertinent question
to ask - some might claim that it is the supreme consideration.
\end{quote}

Now, within the Frequentist framework, the answer is rather disappointing. All we know is that ``the interval $[\ldots]$ covers the true value of $\mu$ with frequency $95\%$
in a long series\label{longrun8} of independent repetitions $[\ldots]$''. Within the Bayesian framework, Barnett's question can be answered, since it assumes a prior distribution on the parameter space and uses ${\bar x}=29.8$ to calculate a posterior distribution about $\mu$.

Because of its crucial dependence on the process generating the data the following phenomenon is also inevitable in the Frequentist framework: Suppose a scientist applies a standard I.Q. test and finds a value of 160 in a certain person. Since the distribution of I.Q. values is known, one can easily give a confidence interval around the observed value. However, suppose ``on the day the score ${\bar x= 160}$ was reported, our test-grading machine was malfunctioning.
Any score ${\bar x}$ below 100 was reported as 100. The machine functioned perfectly
for scores ${\bar x}$ above 100'' \citep[236f]{ef78}. Although the observed value is well above the area where recording errors occured, this new bit of information on the process of data generation alters the confidence interval. Efron concludes: ``it is disturbing that any change at all is necessary. [We received] no new information about the score actually reported, or about I.Q.’s in general. It only concerned something bad that might have happened but didn’t.'' However, he adds: ``Bayesian methods are free from
this defect; the inferences they produce depend only on the data value $\bar x$ actually
observed, since Bayesian averages $[\ldots]$ are conditional on the observed
$\bar x$.''

It is also quite typical that the Frequentist school needs to reframe straightforward questions.
Instead of answering them directly, it considers a similar situation or creates a suitable concept within its own frame of mind.
%That is, it does not answer some question directly. Rather, within its own frame of mind, it considers a similar situation or creates a suitable concept.
In Barnett's example an orthodox statistician would almost surely bring up the prominent notion of unbiased estimation. In Frequentist terms an estimator is a function of the data and some estimator $U$ of $\mu$ is called unbiased if $EU=\mu$. %Since $\bar X$ is unbiased for $\mu$, it does not systematically over- or underestimate the %true but unknown parameter value $\mu$.
How, then, can \citet[332]{pe09} complain that ``$[\ldots]$ one would be extremely hard pressed to find a statistics textbook $[\ldots]$ containing a mathematical proof that randomization indeed produces unbiased estimates of the quantities we wish estimated $[\ldots]$''?

The reason is not hard to find: To him, ``unbiased'' means that all kinds of systematic error are excluded. Quite similarly, the centre for evidence-based medicine at the university of Oxford defines ``Bias: Any tendency to influence the results of a trial (or their interpretation) other than the experimental intervention.'' [Footnote: See www.cebm.net/glossary.] Of course, such a claim is much more difficult to prove than $E{\bar X}=\mu$. It is quite telling that Pearl needs a whole section (6) to illuminate these issues within his elaborated formal framework of causal graphs, while \citet{ja03} has to denote a section (17.2) to the well-known defects of the rather crude classical concept, championed by Neyman. [Footnote: For example, if $EV=\sigma$, i.e. if $V$ is an unbiased estimator of $\sigma$ in the restricted traditional sense, $V^2$ is a biased estimator of $\sigma^2$ (and vice versa). Therefore, early on, \citet[146]{fi73} wished that these considerations would ``have eliminated such criteria as the estimate should be `unbiased' $[\ldots]$'']

 He further explains that ``orthodoxians put such exaggerated emphasis on bias'' due to a ``a psychosematic trap of their own making. When we call the quantity $[EU-\mu]$ the `bias', that makes it sound like something awfully reprehensible, which we must get rid of at all costs $[\ldots]$ Frequentist statisticians adopted the simple device of inventing virtuous-sounding terms (like unbiased, efficient, uniformly most powerful, admissible, robust) to describe their own procedures $[\ldots]$'' \citep[508, 514]{ja03}

Similarly, ``validity'' seems to be exactly what we need. However, randomization only guarantees that a test of significance is valid in a rather narrow, technical sense (for more details see the next section). Overlooking the fact that here, ``valid'' is associated with a restricted meaning, we are deceiving ourselves. Even more so, since randomisation provokes imbalances, thus evoking alternative explanations that threaten internal validity. Jaynes (ibid.) concludes: ``This is just the price one pays for choosing a technical terminology that carries an emotional load, implying value judgements; orthodoxy falls constantly into this tactical error $[\ldots]$ Today these emotionally loaded terms are only retarding progress and doing a disservice to science.''

Hence, despite a ``valid'' framework and mathematically sound conclusions a (purely) Frequentist train of thought may easily miss its target or might even go astray. After decades of Frequentist - Bayesian comparisons like the above, it has become obvious that in many important situations the numerical results of Frequentist and Bayesian arguments (almost) coincide. However, the two approaches are conceptually completely different, and it also has become apparent that simple calculations within the sampling framework lead to reasonable answers to post-data questions only because of ``lucky'' coincidences (e.g. the existence of sufficient statistics for the normal distribution). Of course, in general, such symmetries do not exist, and pre-data results cannot be transferred to post-data situations. In particular, purely Frequentist arguments fail if the sampling distribution does not belong to the ``exponential family'', if there are several nuisance parameters, if there is important prior information, or if the number of parameters is much larger than the number of observations $(p \gg n)$.

It is also no coincidence, but sheer necessity, that a narrow formal line of argument needs to be supplemented with much intuition and heuristics. So, on the one hand, an orthodox author may claim that ``randomization, instrumental variables, and so forth have {\it clear} statistical definitions''; yet, on the other hand, he has to concede at once that ``there is a long tradition of {\it informal} - but systematic and successful - causal inference in the medical sciences'' (see \citet[387]{pe09}, my emphasis).

\section{Random allocation}

In the Frequentist vein
 \begin{quote}
randomization in design $[\ldots]$ is supposed to provide the grounds for {\it replacing} uncertainty about the possible effects of nuisance factors with a probability statement about error'' (\citet[214]{se79}, my emphasis).
\end{quote}
In the light of the above it is straightforward to ask if this ``reframing strategy'' hits its target. Of course, it should come as no surprise that many, if not most, Bayesians have questioned this. For example, towards the end of his article \citet[582]{ba80} writes quite categorically: ``The randomization exercise cannot generate any information on its own. The outcome of the exercise is an ancillary statistic. Fisher advised us to hold the ancillary statistic fixed, did he not?'' Basing our inferences on the distribution randomization creates seems to be the very reverse.

 More recently, philosophers closer to the Bayesian persuation have gained ground \citep{ho06}, and \citet{wo07} explained ``why there is no cause to randomize.'' Yet even by the 1970s, members of the classical school noted that, upon using randomization and the distribution it entails, we are dealing with ``the simplest hypothesis, that our treatment $[\ldots]$ has absolutely no effect in any instance'', and that ``under this {\it very tight} hypothesis this calculation is obviously logically sound'' \citep[my emphasis]{br78}.
Here is a similar, contemporary criticism from an outstanding scientist:
\begin{quote} $[\ldots]$ even under ideal conditions, unaided randomization
cannot answer some very basic questions such as what fraction of a
population benefits from a program $[\ldots]$ Randomization is not an effective procedure for identifying median gains, or the distribution of gains, under general conditions $[\ldots]$ By focusing exclusively on mean outcomes, the statistical
literature converts a metaphor for outcome selection - randomization
- into an ideal $[\ldots]$ \citep[146, 21, emphasis in the original]{he05}.
\end{quote}

In more general terms he (pp. 48, 145, 86) complains:
\begin{quote}
The absence of explicit models is
a prominent feature of the statistical treatment effect literature ``$[\ldots]$ a
large statistical community implicitly appeal to a variety of conventions rather
than presenting rigorous models and assumptions $[\ldots]$ Statistical causal models, in their current state, are
not fully articulated models. Crucial assumptions about sources of randomness
are kept implicit.
\end{quote}

As for the sources of randomness, one should at least distinguish between natural variation and artificially introduced variability. A straightforward question then surely is, how inferences based on the ``man-made'' portion bear on the ``natural'' part. To this end, \citet[579ff]{ba80} compares a scientist, following the logic we described in Section 1 and a statistician who counts on randomization. Let us eavesdrop on their conversation:

\begin{quote}
{\it Statistician:} Observe that the randomization test argument does not depend on any probabilistic assumptions. The randomization probabilities are fully understood and completely under control.

{\it Scientist:} I do not understand the relevance of the randomization probabilities $[\ldots]$ It is relevant to know that the 30 animals have been paired into 15 homogeneous blocks. The manner of my labeling the two animals in the $i$th block $[\ldots]$ does not seem to be of much relevance. The number $m$ of treatment allocations of the type $(t,c)$ seems to be of no consequence at all. $[\ldots]$ How can the level of significance depend so largely on such an irrelevant data characteristic as $m$? $[\ldots]$ I have not been asked on all the background information that I have on the problem $[\ldots]$ I am amazed to find that a statistical analysis of my data can be made without reference to these relevant bits of information.

{\it Statistician:} You are trying to make a joke out of an excellent statistical method of proven value $[\ldots]$ Your criticisms are based on an extreme example and then on a misunderstanding of the very nature of the tests of significance. Tests of significance do not lead to probabilities of hypotheses $[\ldots]$ The randomization analysis of data is so simple, so free of unnecessary assumptions that I fail to understand how anyone can raise any objection against the method.

{\it Scientist:} As a scientist I have been trained to put as much control into the experimental setup as I am capable of, to balance out nuisance factors as far as possible $[\ldots]$ I worked very hard on the project of striking a perfect balance between the treatment and the control groups.

{\it Statistician:} Now the reference consists of only [two points] and the significance level [works out to be $1/2$ or $1$]. Your data is not significant at all.
Had I known about this before, I would not have touched your data with a long pole.

{\it Scientist} (utterly flabbergasted): But my experiment was better planned than a fully randomized experiment, was it not? With my group control (in addition to the usual local control) I made it much harder for $T=\sum t_i -\sum c_i$ to be large in the absence of any treatment difference.

{\it Statistician:} My good man, you must realize that your experiment is no good $[\ldots]$ It appears that you do not have a clear understanding of the role of randomization in statistical experiments.
\end{quote}

Not quite surprisingly, it turns out that the scientist and the statistician are talking past each other. While the foremost goal of the scientist is to make the groups comparable, the statistician focuses on the randomization distribution. Moreover, the scientist asks repeatedly to include important information, but with his inquiry falling on deaf ears, he disputes this statistician's analysis altogether.

\vspace{1ex}
Frequentists say that the crucial role of randomization, stated right at the beginning of this discussion (see the last section), is to provide a known distribution. But is this really so? If the result of a random allocation is extreme (e.g. all women are assigned to T, and all men to C), everybody - Fisher included - seems to be prepared to dismiss this realization: ``It should in fairness be mentioned that, when randomization leads to a bad-looking experiment or sample, Fisher said that the experimenter should, with discretion and judgement, put the sample aside and draw another'' \citep[464]{sa76}.

The latter concession isn't just a minor inconvenience that may be fixed by a ``gentlemen's agreement''. First, an informal correction is wide open to personal capriciousness: (already) ``bad-looking'' to you may be (still) ``fine-looking'' to me. Second, what's the randomization distribution actually used when dismissing some samples? A vague selection procedure will inevitably lead to a badly defined distribution. Third, why reject certain samples at all? If the crucial feature of randomization is to provide a distribution (on which all further ``valid'' inference is based), one should not give away this advantage unhesitantly. At the very least, it is inconsistent to praise the argument of the known framework in theoretical work, and to turn a blind eye to it in practice.

As a matter of fact, in applications, the exact permutation distribution created by some particular randomization process plays a rather subordinate role. Much more frequently, randomization is used as a rationale for common statistical procedures. Here is one of these heuristics: Randomization guarantees independence and if many small uncorrelated (and also often unknown) factors contribute to the distribution of some observable variable $X$, this distribution should be normal - at least approximately, if $n$ is not too small. Therefore, in a statistical experiment, it seems to be justified to compare ${\bar X}_T$ and ${\bar X}_C$, using these means and the sample variance as estimators of their corresponding population parameters - which is nothing but a verbal description of the t-test \citep{go08}. Thus, this test's rationale is supported by randomization. However, it may be noted that Student's famous test was introduced much earlier and worked quite well without randomization's assistance.

Let us look at this from a different angle. A statistical test - or any analytical procedure for that matter - is an algorithm, transferring some numerical input into a certain output which, in the simplest case, is just a single number. From a Frequentist point of view, there are two very different kinds of input: experimental and non-experimental data. However, from a look at the data at hand one cannot tell if they stem from a proper statistical experiment or not. [Footnote: That's the main reason why \citet{pe09} classifies randomization as a causal and not as a statistical concept.] The formal test does not distinguish either: Given the same data it yields exactly the same output. The crucial difference lies in the interpretation of the numerical result. Since randomization treated all variables (known and unknown) alike, the analytical procedure ``catches'' them all and their effect shows up in the output. For example, a confidence interval, so the story goes, gives a quantitative estimate of all of the variables' impact. One can thus numerically assess how strong this influence is, and has, in a sense, achieved explicit quantitative control. In particular, if the combined influence of all nuisance factors (Seidenfeld's ``probability statement about error`'') is numerically small, one may safely conclude that a substantial difference between $T$ and $C$ must be due to the experimental intervention. In a nutshell, owing to randomization, a statistical experiment gives a ``valid'' result in the sense that it allows for far-reaching, in particular causal, conclusions. Seen this way, randomization is sufficient for a causal conclusion, and some are convinced that it is also necessary (Holland's ``no causation without manipulation'', see Section \ref{history}).

Note, however, that the crucial part of the above argument is informal in a rather principled way. It is suspicious that in an experimental, as well as in a similar non-experimental situation, the formal machinery, i.e. the data at hand, the explicit analytical procedure (e.g. a t-test), and the final numerical result may be identical. It is just the narrative prior to the data that makes such a tremendous difference in the end. Since verbal, non-mathematical arguments have a certain power of persuasion which is certainly weaker than a crisp formal derivation or a strict mathematical proof, it seems to be no coincidence that opinion on this matter has remained divided. Followers of Fisher believed in his intuition and trusted randomization, critics did not. And since, sociologically speaking, the Frequentist school dominated the field for decades, so did randomization.

Today, informal reasoning is a bit out of fashion. First, at least in the natural sciences, mathematical arguments are more important than verbal considerations. Typically, the thrust of an argument consists of formulas and their implications, with words of explanation surrounding the formal core. Second, we have learnt that seemingly very convincing verbal arguments can be wrong. In particular, increased formal precision has often corrected intuition and ``obvious'' time-honored conclusions. Hence, the nucleus of most contemporary theories has become mathematical. Causality is no exception to that rule. First, in the last twenty years or so causal networks and causal calculus have formalized this field. Second, as was to be expected, this increased precision straightforwardly demonstrated that certain ``reasonable'' conventions do not work as expected (see \citet{pe09}, in particular Chapter 6 and p. 341). It is not necessary to delve deeply into these matters, since it suffices to look at Savage's example again. He claims qualitatively, and the above models corroborate this quantitatively, that no matter how one splits a heterogenous group into two, the latter groups will always be systematically different. Randomization does not help: If you assign randomly and detect a large effect in the end, your experimental intervention {\it or} the initial difference between $T$ and $C$ may have caused it. All ``valid'' inferential statistics cannot exclude the straightforward second explanation; it's the initial comparability of the groups that is decisive for a sound causal conclusion.

The phrase ``if $n$ is not too small'' is also a verbal argument, implicitly appealing to the central limit theorem. The same with groups created by random assignment: Owing to the latter theorem they tend to become similar. The informal assurance, affirming that this happens fast, ranks among the most prominent conventions of traditional statistics. However, explicit numerical models, e.g. those presented in this contribution, underline that our intuition needs to be corrected. Precise formal arguments - in particular rather straightforward calculations - show that fluctuations cannot be dismissed easily, even if $n$ is large.

Apart from the rather explicit rhetoric of a ``valid framework'', there is also always the implicit logic of the experiment around. Thus, although the received theory emphasizes that ``actual balance has nothing to do with validity of statistical inference; it is an issue of efficiency only'' \citep[227]{se94}; in practice, comparability has turned out to be crucial. Many, if not most, of those praising randomization hurry to mention that it promotes similar groups (see numerous quotations in this article). Nowadays, only a small minority is basing its inferences on the known permutation distribution created by the process of randomization; but an overwhelming majority is checking for comparability. Reviewers of experimental studies routinely request that authors provide randomization checks, that is, statistical tests designed to substantiate the equivalence of $T$ and $C$. At least, in almost every article a list of covariates - with their groupwise means and standard errors - can be found. [Footnote: It is also not much of a surprise that the popular ``propensity score'' is defined as the coarsest {\it balancing} score \citep{ro83}.]

In a nutshell, hardly anybody follows ``pure Frequentist logic''. In a strict sense, there is no logic at all, rather a certain kind of mathematical reasoning plus - since the formal framework is restricted to sampling - a fairly large set of conventions; rigid ``pure'' arguments being readily complemented by applied ``flexibility.'' (The latter consisting of time-honored informal reasoning and shibboleth, but also outright concessions.) Consequently, one finds a broad range of verbal arguments why randomization should be employed:
\begin{quote}
In my view Fisher regarded randomization as being essential in all experiments in the same way that he regarded it as being essential in telepathic and psychophysical experiments: the estimate of error followed exactly from the richness of the randomization. \citep[220]{se94}

As I see it, the purpose of randomization in the design of agricultural field experiments was to help ensure the validity of normal-theory analysis. \citep[583]{hi80}

Randomization tends to produce study groups comparable with respect to known and unknown risk factors, removes investigator bias in the allocation of participants, and guarantees that statistical tests will have valid significance levels. \citep[61]{fr98}

The first property of randomization is that it promotes comparability among the study groups $[\ldots]$ The second property is that the act of randomization provides a probabilistic basis for an inference from the observed results when considered in reference to all possible results.
\citep[7]{ro02a}
\end{quote}

\section{Cause and effect}\label{cause}

The core question of a statistical experiment, in particular a clinical trial, is thus: Does the treatment under consideration work? More specifically: Did the experimental intervention (e.g. a new drug) cause an observed effect (e.g. that the patients in group $T$ lived longer than those in $C$)? The answer will be a straightforward yes, if alternative explanations can be excluded, i.e. if any competing cause can be barred.

In the above chapters we considered two lines of reasoning, the first one going back to J.S. Mill, the second one originating from R.A. Fisher's work. In a sense, they are not competitors. However, at the end of the day, they both should be judged according to their contribution in establishing the desired causal connection.

The logic we have exposed in Section 1 is simple and strong: If the groups are comparable at the beginning and if conscious experimentation guarantees that the intervention remains the only systematic difference between the groups, then this intervention must be the reason for a finally observed difference between the groups.

However, many statisticians count on randomization. They give a variety of reasons why this technique should be used (for a sample see the end of the last section). When they refer to Mill's logic, it is comparability that matters. Alas, as many have noted qualitatively (see Section \ref{randcomp}), and as the above models demonstrate quantitatively, randomization often fails in this respect. Therefore, at least in theoretical discussions, traditional statisticians refer to the ``known-distribution argument'' which, as we have seen in the last section, is also rather short-legged. Finally, there is the ``little-assumption argument:''

  \begin{quote} A new eye drug was tested against an old one on 10 subjects. The
  drugs were randomly assigned to both eyes of each person. In all
  cases the new drug performed better than the old drug. The P-value
  from the observed data is $2^{-10}=0.001$, showing that what we
  observe is not likely due to chance alone, or that it is very
  likely the new drug is better than the old one $[\ldots]$
  Such simplicity is difficult to beat. Given that a physical
  randomization\label{rand1} was actually used, very little extra assumption is
  needed to produce a valid conclusion. \citep[13]{pa01}\label{paw2}\end{quote}

 Since ``there's no such thing as a free lunch'' (Tukey, see \citet[195]{da08}), we should become suspicious upon reading such textbook examples. A narrow, restricted framework is only able to support weak conclusions. Therefore, upon reaching a strong conclusion, one should immediately think about implicit, hidden assumptions. Be reminded of \citet{he05}\label{heck10} who says (pp. 139, 155, his emphasis):

\begin{quote} Structural models do not `make strong assumptions.' They make
explicit the assumptions required to identify
parameters\label{param50} in any particular problem. The treatment
effect literature does not make fewer assumptions; it is much less
explicit about its assumptions $[\ldots]$ The assumptions to justify randomization (no
randomization bias, no contamination or crossover effects
$[\ldots]$) are {\it different} and not weaker or stronger than the
assumptions [econometric models use].

[Footnote: In a table, \citet[87]{he05} compares econometric and statistical causal models. Not surprisingly, the ``range of questions answered'' by the latter is just ``one focused treatment effect.'']
\end{quote}

A second look at Pawitan's example thus reveals that it is the hidden assumption of comparability that carries much of the burden of evidence. It is no coincidence that an eye drug was tested. Suppose, one had tested a liver drug instead: The same numerical result would be almost as convincing if such a drug had been applied to twins or (very) similar persons. However, if the liver drug had been administered to a heterogenous set of persons or if the new drug had been given to a different biological species (mice instead of men, say), exactly the same formal result would not be convincing at all; since, rather obviously, a certain discrepancy a priori may cause a remarkable difference a posteriori.

Comparability keeps this crucial alternative explanation at bay, not randomization; and it is our endeavour to achieve similar groups. Remember that for exactly the same reason Savage (in Section \ref{dicho}) came to the conclusion that ``it
    does not seem possible to base a meaningful experiment on a small heterogenous
    group.'' \citet[52, 70, emphasis in the original]{wa93} adds:
\begin{quote} In statistics, the purpose
of randomization is to achieve homogeneity in the sample units $[\ldots]$ it should
be spelled out that stability and homogeneity are the foundation of the statistical
solution, {\it not} the other way around. For instance, in a clinical trial,
applications of a randomized study to new patients rely on both the stability and
homogeneity assumptions of our biological systems.
\end{quote}

Given this point of view, it turns out that minimization (see Section \ref{compare}) is not just some supplementary technique to improve efficiency. Rather, it is a straightforward and elaborate device to enhance comparability, i.e. to consciously construct similar groups. [Footnote: For the influence of unknown factors see Section \ref{intermediate}.]

\section{A broader perspective}\label{history}

In the 20th century, R.A. Fisher (1890-1962) was the most prominent statistician. However, while his early work on mathematical statistics is highly respected in all quarters, hardly anybody relies on his later ideas, in particular fiducial inference \citep{sa76}. Randomization lies in between, and, quite fittingly, public opinion on this formal technique has been divided. Like the Bayesians, \citet{fi73} was looking for a general inductive logic. However, ``Fisher's main tactic was to logically reduce a given inference problem $[\ldots]$ to a simple form where everyone should agree that the answer is obvious $[\ldots]$ Fisher's inductive logic might be called a theory of types, in which problems are reduced to a small catalogue of obvious situations \citep[97]{ef98}.''

Some fifty years after his death it can safely be said that this approach did not succeed. It turned out that a few remarkable concepts, united in a rather narrow mathematical framework, and augmented by a set of conventions could not reduce the complexity of the real world to a few prototype examples. Leaving out a crucial piece of probability theory, Fisher's strategy was able to give a collection of ad hoc devices, working on many special occasions; but as a whole ``Frequentist theory is shot full\label{shotfull} of contradictions $[\ldots]$'' \citep{ef01}.
  \citet{ba88} and others who doggedly tried to rectify Fisher's ideas, soon became Bayesians. \citet[494ff]{ja03} explains why:
\begin{quote} $[\ldots]$ Jeffreys\label{jeff30} was able to
bypass Fisher's\label{fis182} calculations and derive those
parameter\label{param65} estimates in a few lines of the most elementary algebra
$[\ldots]$ Fisher's\label{fis183} difficult calculations calling for all that
space intuition $[\ldots]$ were quite unnecessary for the actual conduct of
inference. $[\ldots]$ Harold \citet{je39}\label{jeff31} was
able to derive all the same results far more easily, by direct use of
probability theory as logic,\label{indlogik2} and this automatically yielded
additional information about the range of validity of the results\label{valgen}
and how to generalize them, that Fisher\label{fis184} never did
obtain.
\end{quote}

What may thus be said about randomization? How does Fisher's construction of a know distribution perform relative to the classical logic of experimentation?

%\begin{enumerate}
 First, most scientists check for comparability, i.e. they follow Mill's argument. Or, as Frequentist statisticians put it: they all seem to have misunderstood randomization.

  Second, within statistics, Bayesians have always disputed the orthodox point of view, and during the last decades the latter has become less popular.

  Third, even within orthodox statistics, there is no consensus about how to analyze randomized data. Nobody, not even Fisher, relies on ``pure randomization,'' in particular the distribution it generates. It is quite striking that only a minority, perfectly in line with the received position, advices not to adjust data at all. \citet[180f, 191]{fr08b} argues thus:
      \begin{quote}
Regression
adjustments are often made to experimental data. Since randomization does not
justify the models, almost anything can happen $[\ldots]$ The reason
for the breakdown is not hard to find: randomization does not
justify the assumptions behind the OLS model $[\ldots]$ The
simulations,\label{simu2} like the analytic results, indicate a wide
range of possible behavior. For instance, adjustment may help or
hurt.

[Footnote: For similar comments see \citet{fr08a}, \citet[340]{pe09}, and \citet[250ff]{bo14}.]
\end{quote}

Yet a majority, like \citet{ro02, ro02a, sh02}, or \citet{tu00}, opts for an ``adjustment of treatment effect for covariates in clinical trials.'' The latter authors explain why (p. 511):
   \begin{quote}
   Covariates that affect the outcome of a disease are often incorporated into the design
and analysis of clinical trials. This serves two main purposes: 1. To improve the credibility
of the trial results by demonstrating that any observed treatment effect is not accounted
for by an imbalance in patient characteristics, and 2. To improve statistical efficiency.
   \end{quote}
 [Footnote: Notice that, again, the authors readily refer to balance, but not to the ``known-distribution argument''.]

%\end{enumerate}

How strong is the evidence produced by a randomized controlled trial (RCT)? Vis-à-vis the rather anecdotal and qualitative research that preceded today's RCTs, the latter surely constituted real progress. Strict design and standardized analysis has raised the level and has fostered consensus among researchers. However, many classical experiments in the natural sciences are deterministic and have an even better reputation. If in doubt, physicists do not randomize, but replicate. \citet[58, my emphasis]{fi36} gave similar advice:

  \begin{quote}
  [\ldots] no one doubts, in practice, that the probability of being
  led to an erroneous conclusion by the chances of sampling only, can, {\it by
  repetition}
  $[\ldots]$ of the sample, be made so small that the reality of the difference must be
  regarded as convincingly demonstrated.
  \end{quote}

Alas, a large number of important biomedical findings (RCTs included) failed this examination and turned out to be non-replicable \citep{io05, pr11, be12}, so that the \citet{nih13} was forced to launch the ``Replication of Key Clinical Trials Initiative'' [Footnote: Also see \citet{na13}]. The same with experimental psychology which has relied on (small) randomized trials for decades. Now, it is lamenting a ``replicability crisis'' that has proved to be so severe that an unprecedented ``reproducibility project'' needed to be launched \citep{ca12, pa12}. [Footnote: For further similar initiatives see http://validation.scienceexchange.com]

The logic of Section \ref{logic} offers a straightforward explanation to this unpleasant state of affairs. Given (almost) equal initial conditions, the same boundary conditions thereafter, and a well-defined experimental intervention, an effect once observed must reoccur. That's how classical experiments work, which reliably and thus repeatedly hit their target. During the experiment, a controlled environment keeps disturbing factors at bay. Thus, if an effect cannot be replicated, the constructional flaw should be looked for at the very beginning of the endeavour. At this point, it is conspicuous that today's studies do not focus explicitly on the crucial idea of comparability. With other issues - possibly rather irrelevant or even misleading - being at least as important, initial imbalances are the rule and not the exception. At the very least, with randomization, the starting point of researcher 2, trying to repeat the result of researcher 1, will always differ from the latter's point of origin. Therefore, if an effect cannot be replicated, this may well be due to the additional variability introduced by randomization, yielding unequal initial conditions, and ``drowning'' the interesting phenomenon in a sea of random fluctuation.
%And if an effect replicates, there are alternative explanations at hand.
%Therefore even in principle an exact replication is impossible.

% \citet{le00}\label{lesa} Lesaffre sagt: ``$[\ldots]$ it
%is my experience that a successful clinical trial is often attributed to a fast
%recruiting system, efficient clinical staff and a well-organized data management
%system, but less to a proper and clever plan of statistical analysis.''
%Lesaffre, E. (2000)
%Kommentar zu Senn 2000 \citet{se00}. {\it The Statistician} {\bf 49(2)}, 169.

Chance has a Janus face. The idea that many (the more the better), small and rather uncorrelated random influences sum up to a ``mild'' distribution originated in the 19th century, culminating in the famous central limit theorem on which much of classical statistics is built. However, this was not the end of the story. Studying complex systems, physicists soon encountered ``wild'' distributions, in particular power laws \citep[104]{so06}. It is well within this newer set of ideas that a single random event may have a major impact that cannot be neglected (e.g. the energy released by a particularly strong earthquake or the area devastated by a single large flood). %Extreme(r) events do happen,
Quite fittingly, Gumbel said (see \citet[339]{er08}):
\begin{quote}
  It seems that rivers know the theory. It only remains to convince the engineers of the validity of this analysis.
\end{quote}
The fact that the process of randomization can produce a major fluctuation, i.e. a pronounced imbalance in a covariate (and thereby between $T$ and $C$), exerting a tremendous influence on the final result of an RCT is in line with this more recent portrait of chance.

Randomization has tremendous prestige in orthodox statistics, downgrading all designs without a random element to  quasi-experiments. One even distinguishes thoroughly between truly random allocation and ``haphazard assignment, that is, a procedure that is not formally random but has no obvious bias'' \citep[302]{sh02}. Honoring thus the classical philosophical distinction between ``deterministic'' and ``random`'', one readily neglects the fact that modern dynamical systems theory sees a continuum of increasing complexity between perfect (deterministic) order and ``randomness [which] can be thought of as an extreme form of chaos'' \citep[340]{el90}. %Moreover, in ergodic systems, (deterministic) time %averages coincide with (probabilistic) space averages.

With the technique, or rather dogma, of randomization at its heart, Fisher's conception of experiments could even develop into a ``cult of the single study'' \citep[262]{ne99}\label{nel5}, and catch phrases highlighting randomization's outstanding role - e.g. ``no causation without manipulation'' \citep{ho86} - became increasingly popular.
However, this determined point of view has also blocked progress and innovative solutions have been developed elsewhere.

In particular, the story of causation may be told as a ``tale of statistical agony'' \citep{pe09}. Econometrist \citet[5, 147]{he05}\label{heck22} adds:
\begin{quote} Blind empiricism unguided by a theoretical framework for interpreting facts leads nowhere $[\ldots]$ Holland claims that there can be no causal effect of gender on
earnings. Why? Because we cannot randomly assign gender. This confused statement
conflates the act of definition of the causal effect $[\ldots]$ with empirical
difficulties in estimating it $[\ldots]$ This type of reasoning is prevalent in statistics $[\ldots]$
Since randomization is used to define
the parameters of interest, this practice sometimes leads to the confusion that
randomization is the only way - or at least the best way - to identify causal
parameters from real data.\end{quote}

 Heckman earned a nobel prize for his contributions in 2000. Epidemiology, following the \citet{us64}, but not \citet{fi59}, made its way to causal graphs. On the one hand, this formalization straightforwardly gave crucial concepts a sound basis, on the other hand quite a few received ``reasonable'' practices turned out to be dubious \citep{pe09}.

In a more positive vein, computer scientist \citet{ri07} has shown how Fisher's finest ideas may be reformulated and extended within a modern, fine-tuned mathematical framework \citep{li08}. In Rissannen's work one finds a logically sound and general unifying theory of hypothesis testing, estimation and modeling; yet there is no link to randomization. Li and Vitányi even include a chapter (5.5) on ``nonprobabilistic statistics''. In this contemporary theory the crucial concept turns out to be Kolmogorov complexity, allowing to express the idea that a regular sequence $\bf r$ (e.g. ``1,0'' repeated 13 times) is less complex than a sequence like ${\bf s}=(1,0,0,1,0,0,1,1,0,1,1,0,0,0,1,1,1,0,1,1,0,1,0,0,0,0)$ in a mathematically strict way. [Footnote: $\bf s$ can be found in \citet[48]{li08}.]
It also turns out that stochastic processes typically produce complex sequences. However, contrary to the fundamental distinction (deterministic vs. random) mentioned above, given a certain sequence like $\bf s$, it is not possible to tell whether the process that generated $\bf s$ was systematic or not. $\bf s$ could be the output of a ``(pseudo-)random number generator'', i.e. a deterministic algorithm designed to produce chaotic output, or of a truly random mechanism. [Footnote: Whatever that is. For example, strictly speaking, coin tossing - being a part of classical physics - is not.]

Hence, interpreting $\bf s$ as a particular assignment of units to groups ($1 \rightarrow T$, and $0 \rightarrow C$, say), the above fundamental distinction between ```haphazard'' and ``random'' assignment processes seems to be largely overdrawn, some might even question if it is relevant at all. However, it is difficult to deny that the (non-)regularity of the concrete mapping matters. Just compare $\bf r$ to $\bf s$: Since $\bf r$ invites a straightforward alternative explanation, most people would prefer $\bf s$. In today's terminology, Fisher could have been looking for maximally complex sequences, i.e. allocations without any regularity. In his time, a simple stochastic process %(independent of any %subject matter consideration),
typically yielding an ``irregular'' sequence was a straightforward and convenient solution to this problem.

 All in all, Fisher's idea of randomization is still alive. However, at large, it looks more like a remarkable solitaire from a heroic past than like the indispensable key to statistic's future. At first sight, its career has been remarkable, but, make no mistake, it is a sign of crisis if a technique attains the status of a doctrine. This gives those questioning it a hard time. Like all heretics in history, they have received an unfair amount of criticism (just look up the reactions to \citet{ho06}, 1st ed. 1986). Not too long ago, it sufficed to combine the evidence of several studies in a systematic way to be met with scorn and derision. (See, for example, \citet{ey78} on meta-analysis.) %Therefore I think it is a safe prediction that those questioning the tabernacle of the cult of the %single study will receive their share of contempt.

\section{Conclusion: Good experimental practice}

In the face of all the issues we have discussed, Fisher's claim that ``[randomization] relieves the experimenter from the anxiety
of considering and estimating the magnitude of the innumerable causes by which
the data may be disturbed'' seems to be close to wishful thinking. In
particular, quantitative arguments and formal theories show that quite the contrary is true. Since random assignment is {\it no} philosopher's stone, almost effortlessly lifting experimental procedures in the medical and social sciences to
the level of classical experiments in the natural sciences, it has lulled many researchers
into a false sense of security, thereby degenerating the most successful information science \citep{ef01} into a mindless ritual \citep{gi04}.
Why has it taken so long to address the question explicitly? A number of
reasons spring to mind:

%and factors that are not affected by a
%certain balancing schema won't be affected by some random scheme neither.

%\citet[512]{le80} wrote:
%\begin{quote}
%A relatively common preliminary step in assessing therapeutic effects has been
%to compare the distribution of prognostic baseline variables between treatment
%groups to assure that the groups are similar. If each variable appears equally
%distributed between groups (that is, differences are not statistically
%significant), the treatment groups are claimed to be comparable. As our
%experiment indicates, univariate comparisons alone do not necessarily reveal
%baseline differences between treatment groups that can affect the assessment of
%treatment effects.
%\end{quote}

\begin{itemize}
  \item Randomization considerably reduces the operating expense of experimentation
   and makes the analysis of data simple, producing results that seem to be
  far-reaching. Although this sounds too good to be true,
  we are rather inclined to accept such a ``favourable'' procedure,
  in particular, when it is put forward by an authority. To
  admit that there is no such thing as a free lunch, that one has to work
  hard in order to get (closer to) comparability, is unpleasant
  news that we tend to avoid.
  \item Randomization addresses the fundamental problem of controlling unknown
  nuisance factors. In particular, it is impartial and does not favour or discriminate against any
  particular variable. This seems to outweigh the obvious disadvantage of randomization,
  i.e., known factors are less balanced than they would be if one used some non-randomized
  procedure, such as minimization.

  \item Randomization fits well into the dominant Frequentist framework. Moreover, as
  the fundamental difference between random quantities and their
  realizations is rather blurred in traditional statistics, one easily confuses the
  two. Yet, as the above models and quotations show, the difference between
  process and realization is crucial: Although
  a random allocation mechanism is independent of any other variable, its effect, a
  particular allocation of units to groups, can be considerably out of balance.

\end{itemize}

  With respect to the last point it is often said that ``randomization equates
  the groups on expectation'' (e.g. \citet[40, 82]{mo07}, \citet[250]{sh02}). The latter
  authors explain: ``In any given experiment, observed pretest means will differ due to
  luck $[\ldots]$ But we can expect that participants will be equal over
  conditions in the long run over many randomized experiments.''

  What is wrong with this argument? First, it conflates the {\it single experiment} being conducted and analyzed with a {\it hypothetical series} of experiments. Second,
  it therefore downplays the fundamental problem pointed out in Section \ref{logic}
  and emphasized by many authors (see Section \ref{randcomp}): In each and every
  experiment there is a convincing alternative explanation if
  the groups differ from the outset. Third, these substantial objections do not just vanish ``in the
  long run''. Rather, if each study can be seriously challenged, the
  whole of the evidence may remain rather shaky. Fourthly, although it is certainly correct that replicated, randomized
  experiments align the set of all those treated with the set of all those not treated
  (``in expectation''), this simply does not imply that such a symmetry property also holds
  in the single experiment. Typically, quite the reverse is true (see Section \ref{models}).
Finally, the argument praises randomization
  where credit ought to be given to replication.

 % We trust a stable result, i.e.,
 % a series of experiments (almost) all pointing in the same direction.
 % Randomization helps in that matter since it is very unlikely that unfavourable realizations of
 % nuisance factors - ``Lady Luck'' - has always been the reason for a permanently
 % observable effect. If the probability of such an artifact is $p$ in a
 % single experiment, $k$ independent replications of this study should reduce the probability of this kind of error %to $p^k$.
 % Up-to-date methodologies such as meta-analysis and systematic reviews that (more and less formally) accrue the %evidence of several
 % experiments are based on the same idea.
 % Moreover, many replicable and highly
 % trustworthy ``deterministic'' experiments demonstrate that a random design element
 % is not crucial.

Coming back to the main theme of this contribution, it may be said that chance in the guise of randomization by and large supports comparability. However,
since the former is blind with respect to concrete factors and relevant interactions that
may be present, it needs a large number of experimental units to do so. The intuition behind this result is easy to grasp: Without
knowledge of the subject-matter, randomization has to protect against every conceivable
nuisance factor. Such a kind of unsystematic protection is provided by number and
builds up slowly. Thus, a huge number of randomly allocated subjects is needed to
shield against a moderate number of potential confounders. And, of course, no finite
procedure such as the flip of a coin is able to control for an {\it infinite} number of
nuisance variables.

Therefore, it seems much more advisable to use background knowledge in order to
minimize the difference between groups with respect to known factors or specific
threats to experimental validity. As of today, minimization operationalizes this idea best. At the end of such a conscious construction
process, randomization finds its proper place. Only if no reliable context
information exists is unrestricted randomization the method of choice. It must be
clear, however, that it is a weak guard against confounding, yet the only one
available in such inconvenient situations.

All in all, the above analysis strongly recommends traditional experimentation, thoroughly
selecting, balancing and controlling factors and subjects with respect to known
relevant variables, thereby using broader context information - i.e., substantial scientific knowledge. I agree with
\citet[76f]{pe03}\label{pen} who says:
%Schon indem man $n$ fixiert geht man von einem
%endlichen Informationsumfang aus, mit dessen Hilfe man eine absicherte
%0-1-Entscheidung treffen will.
\begin{quote}
$[\ldots]$ it is the existence of sound background theory which is crucial for the
success of science. It is the framework against which observations are made, it
allows strict definition of the items involved, it is the source of information
about possible relevant variables and allows for the identification of homogeneous
reference classes that ensure regularity and, hence, reliable causal inference.
\end{quote}

Cumulative science is the result of a successful series of such experiments - each of them focusing on the crucial ingredients, like precise research questions, convincing operationalizations, explicit control, quantitative measures of effect, and - last but not least - comparability.

\end{doublespace}


\begin{thebibliography}{99}

\bibitem[\protect\citeauthoryear {Altman}{1985}]{al85} Altman, D. G. (1985).
Comparability of Randomised Groups. {\it The Statistician} {\bf 34}, 125-136.

\bibitem[\protect\citeauthoryear {Austin}{2008}]{au08} Austin, P. C. (2008).
A Critical Appraisal of Propensity-Score Matching in the Medical Literature
between 1996 and 2003. {\it Statistics in Medicine} {\bf 27}, 2037-2049.

%\bibitem[\protect\citeauthoryear {Bailey}{1982}]{ba82} Bailey, R. A.
%(1982). Randomization, Constrained. In: Johnson, N. L,; and Kotz, S. (eds.)
%Encyclopedia of Statistical Sciences {\it Wiley, New York.}

\bibitem[\protect\citeauthoryear {Barnard}{1993}]{ba93} Barnard, G. A.
(1993). Discussion of Draper, D.; Hodges, J. S.; Mallows, C. L.; and Pregibon,
D. (1993). Exchangeability and Data Analysis. {\it J. Royal Stat. Soc. A} {\bf
56(1)}, 9-37.

\bibitem[\protect\citeauthoryear {Barnett}{1999}]{ba99} Barnett, V. (1999).
Comparative Statistical Inference. (3. ed.)  {\it Wiley, New York}. 1. ed. 1973.

\bibitem[\protect\citeauthoryear {Basu}{1980}]{ba80} Basu, D. (1980).
Randomization Analysis of Experimental Data: the Fisher Randomization Test. {\it J.
of the American Statistical Association} {\bf 75}, 575-595.

\bibitem[\protect\citeauthoryear {Basu}{1988}]{ba88} Basu, D. (1988).
Statistical Information and Likelihood. A Collection of Critical Essays by Dr. D.
Basu, ed. by J.K Gosh. {\it Springer, New York: Lecture Notes in Statistics}.

\bibitem[\protect\citeauthoryear {Begley and Ellis}{2012}]{be12}
Begley, C.G.; and Ellis, L.M. (2012). Raise standards for preclinical cancer research. {\it Nature} {\bf 483}, 531-533.

\bibitem[\protect\citeauthoryear {Berger}{2005}]{be05} Berger, V. (2005).
Selection Bias and Covariate Imbalances in Randomized Clinical Trials. {\it
Wiley, New York}.

%\bibitem[\protect\citeauthoryear {Berger und Weinstein}{2004}]{be04} Berger, V.;
%and Weinstein, S. (2004). Ensuring the Comparability of Comparision Groups: Is
%Randomization Enough? {\it Controlled Clinical Trials} {\bf 25}, 515-524.

\bibitem[\protect\citeauthoryear {Bookstein}{2014}]{bo14} Bookstein, F. L. (2014).
Measuring and Reasoning. Numerical Inference in the Sciences. {\it Cambridge University Press, New York}.

%\bibitem[\protect\citeauthoryear {Boring}{1919}]{bo19} Boring, E. G. (1919).
%Mathematical vs. Scientific Significance. {\it Psychological Bulletin} {\bf
%16(10)}, 335-338.

\bibitem[\protect\citeauthoryear {Boring}{1953}]{bo53} Boring, E. G. (1953).
The Nature and History of Experimental Control. {\it The American Journal of
Psychology} {\bf 67(4)}, 573-589.

\bibitem[\protect\citeauthoryear {Box et al.}{2005}]{bo05} Box, G. E. P.; Hunter, J. S.;
and Hunter, W. G. (2005). Statistics for Experimenters. Design, Innovation, and
Discovery. (2. ed.) {\it Wiley, New York}.

\bibitem[\protect\citeauthoryear {Brillinger et al.}{1978}]{br78}
Brillinger, D.R.; Jones; L.V.; and Tukey, J.W. (1978). The Role of Statistics in
Weather Resources Management. Report of the Statistical Task Force to the
Weather Modification Advisory Board. {\it Government Printing Office, Washington
D. C}.

\bibitem[\protect\citeauthoryear {Carpenter}{2012}]{ca12}
Carpenter, S. (2012). Psychology’s bold initiative. {\it Science}, {\bf 335}, 1558–1560.

\bibitem[\protect\citeauthoryear {Chu et al.}{2012}]{ch12} Chu, R.; Walter, S.D.; Guyatt, G.; Devereaux, P.J.; Walsh, M.; Thorlund, K.; and Thabane, L. (2012).
Assessment and Implication of Prognostic Imbalance in
Randomized Controlled Trials with a Binary Outcome – A
Simulation Study. {\it PLOS One} {\bf 7(5), e36677}.

\bibitem[\protect\citeauthoryear {Cohen}{1988}]{co88} Cohen, J. (1988).
Statistical Power Analysis for the Social Sciences. (2. ed.) {\it Erlbaum,
Hillsdale, NJ.} 1. ed. 1969.

\bibitem[\protect\citeauthoryear {Cornfield}{1976}]{co76} Cornfield, J. (1976).
Recent Methodological Contributions to Clinical Trials. {\it American Journal of
Epidemiology} {\bf 104(4)}, 408-421.

\bibitem[\protect\citeauthoryear {Davies}{2008}]{da08}
Davies, P.L. (2008). Approximating Data (with discussion). {\it J. of the Korean
Statistical Society} {\bf 37}, 191-240.

\bibitem[\protect\citeauthoryear {de Finetti}{1974}]{fi74} Finetti, B. de (1974).
Theory of Probability. {\it Wiley, London}.

\bibitem[\protect\citeauthoryear {Efron}{1978}]{ef78} Efron, B. (1978).
Controversies in the Foundations of Statistics. {\it American Math. Monthly}
{\bf 85(4)}, 232-246.

\bibitem[\protect\citeauthoryear {Efron}{1998}]{ef98} Efron, B. (1998).
R. A. Fisher in the 21st Century (with discussion). {\it Statistical Science} {\bf
13(2)}, 95-122.

\bibitem[\protect\citeauthoryear {Efron}{2001}]{ef01} Efron, B. (2001).
Statistics is the Science of Information Gathering, Especially when
the Information arrives in Little Pieces instead of Big Ones. Interview with Bradley
Efron. Business Science Center, Irwin/McGraw-Hill Student Learning Aids.
www.mhhe.com/business/opsci/bstat/efron.mhtml

\bibitem[\protect\citeauthoryear {Elston und Glasbey}{1990}]{el90}
Elston, D.A.; and Glasbey, C.A. (1990). {\it J. of
the Royal Statistical Society, Ser. A} {\bf 153(3)}, 340-341. Comment on Bartlett, M.S.
Chance or Chaos?, 321-347.

\bibitem[\protect\citeauthoryear {Érdi}{2008}]{er08} Érdi, P. (2008).
Complexity explained. {\it Springer, Berlin and Heidelberg}.

\bibitem[\protect\citeauthoryear {Eysenck}{1978}]{ey78} Eysenck, H.J. (1978).
An exercise in mega-silliness. {\it American Psychologist} {\bf
33(5)}, 517.

\bibitem[\protect\citeauthoryear {Fisher}{1935}]{fi35} Fisher, R. A. (1935).
The Design of Experiments. Cited according to the 8th edition (1966) by {\it
Hafner Publishing Company, New York}.

\bibitem[\protect\citeauthoryear {Fisher}{1936}]{fi36} Fisher, R.A. (1936).
``The Co-efficient of Racial Likeness'' and the Future of Craniometry. {\it J.
of the Royal Anthropological Institute} {\bf 66}, 57-63.

\bibitem[\protect\citeauthoryear {Fisher}{1959}]{fi59} Fisher, R.A. (1959).
Smoking: the Cancer Controversy. {\it Oliver and Boyd, Edinburgh}.

\bibitem[\protect\citeauthoryear {Fisher}{1973}]{fi73} Fisher, R.A. (1973).
Statistical Methods and Scientific Inference. (3. ed) {\it Hafner
Publishing Company, New York}.

\bibitem[\protect\citeauthoryear {Freedman}{2008a}]{fr08a} Freedman, D. A. (2008a).
Randomization does not Justify Logistic Regression. {\it Statistical Science}
{\bf 23}, 237-249.

\bibitem[\protect\citeauthoryear {Freedman}{2008b}]{fr08b} Freedman, D. A. (2008b).
On Regression Adjustments to Experimental Data. {\it Advances in Applied
Mathematics} {\bf 40}, 180-193.

%\bibitem[\protect\citeauthoryear {Freedman}{2010}]{fr10} Freedman, D.A. (2010).
%Statistical Models and Causal Inference. A Dialogue with the Social Sciences.
%Edited by Collier, D.; Sekhon, J.S.; and Stark, P.B. {\it Cambridge University
%Press, New York}.

\bibitem[\protect\citeauthoryear {Friedman et al.}{1998}]{fr98} Friedman, L. M.;
Furberg, C.D.; and DeMets, D.L. (1998). Fundamentals of Clinical Trials. (3.
ed.) {\it Springer}.

\bibitem[\protect\citeauthoryear {Gigerenzer}{2004}]{gi04} Gigerenzer, G. (2004).
Mindless Statistics. {\it The Journal of Socio-Economics} {\bf 33}, 587-606.

%\bibitem[\protect\citeauthoryear {Goodman}{1999}]{go99} Goodman, S. N. (1999).
%Toward Evidence-Based Medical Statistics. 1: The $p$ Value Fallacy. {\it Annals
%Intern Med.} {\bf 130}, 995-1004.

\bibitem[\protect\citeauthoryear {Gosset}{1908}]{go08} Gosset, W. (1908).
The Probable Error of a Mean. {\it Biometrika} {\bf 6(1)}, 1-25.

\bibitem[\protect\citeauthoryear {Greenland}{1990}]{gr90} Greenland, S. (1990).
Randomization, Statistics, and Causal Inference. {\it Epidemiology} {\bf 1(6)},
421-429.

\bibitem[\protect\citeauthoryear {Greenland et al.}{1999}]{gr99} Greenland, S.;
Robins, J.M.; and Pearl, J. (1999). Confounding and Collapsibility in Causal
Inference. {\it Statistical Science} {\bf 14(1)}, 29-46.

%\bibitem[\protect\citeauthoryear {Guttman}{1985}]{gu85} Guttman, L. (1985).
%The Illogic of Statistical Inference for Cumulative Science. {\it Applied
%stochastic models and data analysis} {\bf 1}, 3-9.

\bibitem[\protect\citeauthoryear {Hacking}{1988}]{ha88} Hacking, I. (1988).
Telepathy: Origins of Randomization in Experimental Design. {\it ISIS} {\bf 79},
427-451.

\bibitem[\protect\citeauthoryear {Heckman}{2005}]{he05} Heckman, J. J. (2005).
The Scientific Model of Causality. {\it Sociological Methodology} {\bf 35},
1-162.

\bibitem[\protect\citeauthoryear {Hinkley}{1980}]{hi80} Hinkley, D. V.
(1980). Remark on \citet{ba80}. {\it J. of the American Statistical
Association} {\bf 75}, 582-584.

\bibitem[\protect\citeauthoryear {Holland}{1986}]{ho86}
Holland, P.W. (1986) Statistics and Causal Inference (with discussion) {\it J.
of the American Statistical Association} {\bf 81}, 945-970.

\bibitem[\protect\citeauthoryear {Howson and Urbach}{2006}]{ho06} Howson, C.; and
Urbach, P. (2006). Scientific Reasoning. The Bayesian Approach. (3. ed.) {\it
Open Court, Chicago and La Salle, IL}.

\bibitem[\protect\citeauthoryear {Ioannidis}{2005}]{io05} Ioannidis, J.P.A. (2005).
Contradicted and initially stronger effects in highly cited clinical research. {\it JAMA} {\bf 294},
218-229.

\bibitem[\protect\citeauthoryear {Jaynes}{2003}]{ja03} Jaynes, E.T. (2003).
Probability Theory. The Logic of Science. {\it Cambridge University Press,
Cambridge}.

\bibitem[\protect\citeauthoryear {Jeffreys}{1939}]{je39} Jeffreys, H. (1939).
Theory of Probability. {\it Clarendon Press, Oxford}.

\bibitem[\protect\citeauthoryear {Johnson et al.}{2005}]{jo05} Johnson, N.L.; Kemp, A.W.;
and Samuels, S. (2005). Univariate Discrete Distributions. (3. ed.) {\it Wiley}.

\bibitem[\protect\citeauthoryear {Johnstone}{1988}]{jo88} Johnstone, D. J.
(1988). Hypothesis Tests and Confidence Intervals in the Single Case. {\it
British J. for the Philosophy of Science} {\bf 39}, 353-360.

\bibitem[\protect\citeauthoryear {Kadane and Seidenfeld}{1990}]{ka90} Kadane, J.B.;
and Seidenfeld, T. (1990). Randomization in a Bayesian Perspective. {\it Journal of
Statistical Planning and Inference} {\bf 25}, 329-345.

%\bibitem[\protect\citeauthoryear {Keiding}{1994}]{ke94} Keiding, N. (1994).
%Commentary on Spielhalter, D. J.; Freedman, L. S.; and Parmar, M. K. B. (1994).
%Bayesian Approaches to Randomized Trials. {\it J. of the Royal Statistical
%Society, Ser. A} {\bf 157 (3)}, 395.

%\bibitem[\protect\citeauthoryear {Kempthorne}{1992}]{ke92} Kempthorne, O.
%(1992). Intervention Experiments, Randomization and Inference. In: Ghosh, M.;
%and Pathak, P. K. (Eds., 1992). Current Issues in Statistical Inference: Essays
%in Honor of D. Basu. {\it Institute of Mathematical Statistics: Lecture Notes -
%Monograph Series}, 13-31.

\bibitem[\protect\citeauthoryear {Lee et al.}{1980}]{le80} Lee, K. L.; MnNeer, J. F.;
Starmer, C. F.; Harris, P. J.; and Rosati, R. A. (1980). Clinical judgment and
statistics. Lessons from a simulated randomized trial in coronary artery
disease. {\it Circulation} {\bf 61}, 508-515.

%\bibitem[\protect\citeauthoryear {Lindley}{1982}]{li82} Lindley, D. (1982). The
%Role of Randomization in Inference. In: Asquith,P. D.; and Nickles, T. (Hrsg.)
%PSA 1982. Proceedings of the 1982 biennial meeting of the philosophy of science
%association, vol. 2 `Symposia'. {\it Philosophy of Science Association, East
%Lansing, MI}, 431-446.

\bibitem[\protect\citeauthoryear {Li und Vitányi}{2008}]{li08} Li, M.; and Vitányi, P.
(2008). An Introduction to Kolmogorov
Complexity and its Applications. (3rd ed.) {\it Springer, New
York}.

\bibitem[\protect\citeauthoryear {Lindley}{1982}]{li82} Lindley, D.V. (1982).
The Role of Randomization in Inference. In: Asquith,P.D.; and Nickles, T.
(eds.) PSA 1982. Proceedings of the 1982 biennial meeting of the philosophy of
science association, vol. 2 ``Symposia''. {\it Philosophy of Science Association,
East Lansing, MI}, 431-446.

\bibitem[\protect\citeauthoryear {Mill}{1843}]{mi43} Mill, J. S. (1843).
A System of Logic, Ratiocinative and Inductive. {\it London}. Cited according to
the 1859 ed. by {\it Harper \& Brothers, New York}.

\bibitem[\protect\citeauthoryear {Morgan und Winship}{2007}]{mo07} Morgan, S.L.;
and Winship, C. (2007). Counterfactuals and Causal Inference. {\it Cambridge
University Press, Cambridge}.

\bibitem[\protect\citeauthoryear {National Institute of Health}{2013}]{nih13}
National Institute of Health (2013). 	
Replication of Key Clinical Trials Initiative. See
https://grants.nih.gov/grants/guide/pa-files/PAR-13-383.html

\bibitem[\protect\citeauthoryear {the editors of Nature}{2013}]{na13}
Nature (2013). Announcement: Reducing our irreproducibility. Editorial. {\it Nature} {\bf 496}, 398. See also:
www.nature.com/nature/focus/reproducibility

\bibitem[\protect\citeauthoryear {Oxford Centre for Evidence-based Medicine}{2009}]{ox09}
Oxford Centre for Evidence-based Medicine (2009). Levels of Evicence. See
www.cebm.net/index.aspx?o=1025

\bibitem[\protect\citeauthoryear {Nelder}{1999}]{ne99}
Nelder, J.A. (1999). Statistics for the Millenium (with discussion). {\it The
Statistician} {\bf 48(2)}, 257-269.

\bibitem[\protect\citeauthoryear {Pashler and Wagenmakers}{2012}]{pa12}
Pashler, H.; and Wagenmakers, E.-J. (2012). Editors’ Introduction to the Special Section on Replicability in Psychological Science A Crisis of Confidence? {\it Perspectives on Psychological Science} {\bf 7(6)}, 528-530.

\bibitem[\protect\citeauthoryear {Pawitan}{2001}]{pa01} Pawitan, Y. (2001).
In all Likelihood: Statistical Modelling and Inference Using
Likelihood. {\it Clarendon Press, Oxford.}

%\bibitem[\protect\citeauthoryear {Pearl}{2000}]{pe00} Pearl, J. (2000).
%Causality. Models, Reasoning and Inference. {\it Cambridge University Press.}

\bibitem[\protect\citeauthoryear {Pearl}{2009}]{pe09} Pearl, J. (2009).
Causality. Models, Reasoning and Inference. (2. ed.) {\it Cambridge University
Press.}

%\bibitem[\protect\citeauthoryear {E. S. Pearson}{1938}]{pe38} Pearson, E.S.
%(1938). Student as Statistician. {\it Biometrika} {\bf 30}, 210-250.

\bibitem[\protect\citeauthoryear {Penston}{2003}]{pe03} Penston, J. (2003).
Fiction and Fantasy in Medical Research. The Large-Scale Randomised Trial. {\it The
London Press, London.}

\bibitem[\protect\citeauthoryear {Prinz et al.}{2011}]{pr11}
Prinz, F.; Schlange, T.; and Asadullah, K. (2011). Believe it or not: how much can we rely on published data on potential drug targets? {\it Nat Rev Drug Discovery} {\bf 10(9)},
712.

\bibitem[\protect\citeauthoryear {Rissanen}{2007}]{ri07} Rissanen, J. (2007).
Information and Complexity in Statistical Modelling. {\it Springer, New York}.

\bibitem[\protect\citeauthoryear {Rosenbaum}{2002}]{ro02} Rosenbaum, P. R. (2002).
Observational Studies. (2. ed.) {\it Springer, New York: Springer Series in
Statistics}. 1. ed. 1995.

\bibitem[\protect\citeauthoryear {Rosenbaum und Rubin}{1983}]{ro83} Rosenbaum, P.R.;
and Rubin, D.B. (1983). The Central Role of the Propensity Score in
Observational Studies for Causal Effects. {\it Biometika}, {\bf 70(1)}, 41-55.

\bibitem[\protect\citeauthoryear {Rosenberger and Lachin}{2002}]{ro02a} Rosenberger, W. F.; and Lachin, J. M. (2002). Randomization in Clinical Trials. Theory and Practice. {\it J. Wiley \& Sons, New York}.

\bibitem[\protect\citeauthoryear {Rubin}{1978}]{ru78} Rubin, D.B. (1978).
Bayesian Inference for Causal Effects: The Role of Randomization. {\it Annals of
Statistics} {\bf 6}, 34-58.

\bibitem[\protect\citeauthoryear {Rubin}{2006}]{ru06} Rubin, D. B. (2006).
Matched Sampling for Causal Effects. {\it Cambridge University Press, Cambridge}.

%\bibitem[\protect\citeauthoryear {Saint-Mont}{2011}]{sa11} Saint-Mont, U. (2011).
%Statistik im Forschungsprozess. [The Role of Statistics in the Research Process.]
%{\it Physica Verlag, Heidelberg}.

%\bibitem[\protect\citeauthoryear {Salsburg}{1985}]{sa85} Salsburg, D. S. (1985).
%The Religion of Statistics as practiced in Medical Journals. {\it The American
%Statistician} {\bf 39(3)}, 220-223.

\bibitem[\protect\citeauthoryear {Savage}{1962}]{sa62} Savage, L.J. (1962).
In: Cox, D.R.; and Barnard, G.A. The foundations of statistical inference. A
discussion. {\it Methuen \& Co LTD, London; and J. Wiley \& Sons, New York}.

\bibitem[\protect\citeauthoryear {Savage}{1976}]{sa76} Savage, L.J.H. (1976).
On Rereading R. A. Fisher (with discussion). {\it Annals of Statistics} {\bf 4(3)},
441-500.

\bibitem[\protect\citeauthoryear {Seidenfeld}{1979}]{se79} Seidenfeld, T. (1979).
Philosophical Problems of Statistical Inference. Learning from R. A. Fisher.
{\it D. Reidel, Dordrecht.}

\bibitem[\protect\citeauthoryear {Senn}{1994}]{se94} Senn, S. (1994).
Fisher's Game with the Devil. {\it Statistics in Medicine} {\bf 13}, 217-230.

\bibitem[\protect\citeauthoryear {Senn}{2000}]{se00} Senn, S. (2000).
Consensus and Controversy in Pharmaceutical Statistics (with discussion). {\it
The Statistician} {\bf 49(2)}, 135-176.

\bibitem[\protect\citeauthoryear {Shadish et al.}{2002}]{sh02} Shadish, W.R., Cook, T.D, and Campbell, D.T. (2002). Experimental
and Quasi-Experimental Designs for Generalized Causal Inference. {\it Houghton
Mifflin Company}.

\bibitem[\protect\citeauthoryear {Simpson}{1951}]{si51}
Simpson, E.H. (1951). The Interpretation of Interaction in Contingency Tables. {\it
J. of the Royal Statistical Society, Ser. B} {\bf 13}, 238-241.

\bibitem[\protect\citeauthoryear {Sornette}{2006}]{so06}
Sornette, D. (2006). Critical phenomena in natural sciences. {\it Springer Series in Synergetics, Berlin}.

\bibitem[\protect\citeauthoryear {Advisory Committee to the Surgeon General of the Public Health Service}{1964}]{us64} U.S.
Department of Health, Education, and Welfare (1964). Smoking and Health: Report of the
Advisory Committee to the Surgeon General of the Public Health Service. {\it
Public Health Service Publication No. 1103}, Washington D.C.

\bibitem[\protect\citeauthoryear {Taves}{1974}]{ta74}
Taves, D.R. (1974). Minimization: A new Method of Assigning Patients to Treatment
and Control Groups. {\it Clinical Pharmacology and Therapeutics} {\bf 15(5)},
443-453.

\bibitem[\protect\citeauthoryear {Treasure and MacRae}{1998}]{tr98}
Treasure, T.; and MacRae, K.D. (1998). Minimisation: the platinum standard for
trials? Randomisation doesn't guarantee similarity of groups; minimisation does.
{\it British Medical J.} {\bf 317}, 362–363.

%\bibitem[\protect\citeauthoryear {Trollope}{1867}]{tr67}
%Trollope, A. (1867). Phineas Finn. {\it eBooks@Adelaide}, Chapter 60.

\bibitem[\protect\citeauthoryear {Tu et al.}{2000}]{tu00}
Tu, D.; Shalay, K.; and Pater, J. (2000). Adjustment of treatment effect for covariates in clinical trials: Statistical and regulatory issues.
{\it Drug Information Journal} {\bf 34}, 511-523.

\bibitem[\protect\citeauthoryear {Urbach}{1985}]{ur85} Urbach, P. (1985).
Randomization and the Design of Experiments. {\it Philosophy of Science} {\bf
52}, 256-273.

\bibitem[\protect\citeauthoryear {Wang}{1993}]{wa93} Wang, C. (1993).
Sense and Nonsense of Statistical Inference. Controversy, Misuse and Subtlety. {\it
Marcel Dekker, New York}.

\bibitem[\protect\citeauthoryear {Worrall}{2007}]{wo07} Worrall, J. (2007).
Why There's No Cause to Randomize. {\it Brit. J. Phil. Sci.} {\bf 58}, 451-488.

\end{thebibliography}
\end{document}